\documentclass[a4paper,fleqn,usenatbib]{mnras}

\usepackage{amsmath}	

\usepackage[T1]{fontenc}
\usepackage{ae,aecompl}

\usepackage{graphicx}	

\usepackage{amssymb}	

\title[Dust across the Universe]{The cosmic dust rate across the Universe}
\author[L. Gioannini et al.]{L. Gioannini$^{1}$\thanks{E-mail:
gioannini@oats.inaf.it}, F. Matteucci$^{1,2,3}$, F. Calura$^4$
\\
$^{1}$Dipartimento di Fisica, Sezione di Astronomia, Universit\`a di Trieste, via G.B. Tiepolo 11, 34100, Trieste, Italy\\
$^{2}$INAF, Osservatorio Astronomico di Trieste, via G.B. Tiepolo 11, 34100, Trieste, Italy\\
$^{3}$INFN, Sezione di Trieste, Via Valerio 2, 34100, Trieste, Italy\\
$^{4}$INAF, Osservatorio Astronomico di Bologna, Via Gobetti 93/3, 40129 Bologna, Italy}

\begin{document}

\date{Accepted for Publication by MNRAS on 2017 July 25.}

\pagerange{\pageref{firstpage}--\pageref{lastpage}} \pubyear{2017}

\maketitle

\label{firstpage}

\begin{abstract}
We investigate the evolution of interstellar dust in the Universe by means of chemical evolution models of galaxies of different morphological types, 
reproducing the main observed features of present day galaxies. We adopt the most updated prescriptions for dust production from 
supernovae and asymptotic giant branch (AGB) stars as well as for dust accretion and destruction processes.
Then, we study the cosmic dust rate in the framework of three different cosmological scenarios for galaxy formation:
i) a pure luminosity scenario (PLE), 
ii) a number density evolution scenario (DE), as suggested by the classical hierarchical clustering scenario and 
iii) an alternative scenario, in which both spirals and ellipticals are allowed to evolve in number on an observationally motivated basis.
Our results give predictions about the evolution of the dust content in different galaxies as well as the cosmic dust rate as a function of redshift.
Concerning the cosmic dust rate, the best scenario is the alternative one, which predicts a peak at $2<z<3$ and reproduces the cosmic star formation rate.
We compute the evolution of the comoving dust density parameter $\Omega_{dust}$ and find agreement with data for $z< 0.5$ in the framework of DE and alternative scenarios.
Finally, the evolution of the average cosmic metallicity is presented and it shows a quite fast increase in each scenario, 
reaching the solar value at the present time, although most of the heavy elements are incorporated into solid grains, and therefore not observable in the gas phase.
\end{abstract}

\begin{keywords}
ISM: dust, extinction -- ISM: abundances --  galaxies: evolution -- galaxies: high redshift -- galaxies: star formation -- ISM: abundances
\end{keywords}

\section{Introduction}
The interstellar dust is a very important constituent of our Universe as it is involved 
in a great variety of physical processes:
it acts as a catalyst for the formation of molecular hydrogen \citep{Ho_Sa71,Ma90}; it
absorbs the infrared (IR) stellar light and re-emits in the ultra-violet (UV) band \citep{Dr_Lee84,De90,Wi_Go00}; it 
deeply affects the spectral energy distribution of galaxies \citep{Silva1998, Granato2000} and 
depletes metals from the gas phase of the interstellar medium (ISM) \citep{Je09,Vl04}. 
Dust is first injected into the ISM through stellar production by asymptotic giant branch stars (AGBs) and Type II Supernovae (SNe) 
\citep{Ha96,Fe_Ga_06,Ga09,Ve12,To_Fe_01,Bi_Sc_07,Bo14}, and,
once in the ISM, it experiences many processes which are able to change both its mass and size of grains. 
Dust growth in the dense ISM increases the dust mass, thanks to 
the metal accretion onto pre-existing dust grains \citep{Li_Cl_89,Dr90,Dr09,Hi_Ku_11}.
On the other hand, dust destruction via thermal sputtering, grain collisions or supernova shocks 
restores the mass to the gas phase \citep{Dr_Sa_79,Dw_Sc_80,Jo94}. 

Dust has been detected in various environments both in the local and in the high redshift Universe. 
During the past years, several studies tried to assess the total amount of dust in the Universe 
at different scales and redshifts \citep{Fu11,Me_Fu12,De_Co12,Cl15}.
Dust properties in galaxies are usually studied by means of various scaling relations, mostly involving
dust, stellar and gas mass \citep{Du03,Santini2014,Po16,Ca17,Cas17}. 
In particular, the relation between the gas-to-dust ratio and the metallicity represents an 
important diagnostic tool to understand dust evolution and formation in galaxies 
\citep{Gala11,Gin17,Re14}, and it can place important constraints on models:
in fact, the dust-to-gas ratio connects the metal mass embedded in the dust phase to the gas phase, 
whereas the metallicity is a fundamental parameter in studying galaxy evolution.

Far from the local Universe, 
Damped Lyman Alpha systems (DLAs) offer a great opportunity to investigate the dust properties 
in the ISM in a redshift range 1<z<5 \citep{Pei91,Pe94,Vla02,Ca03,Vl_Pe05,Dec16}.
More distant, high redshift galaxies (z<5) host a significant reservoir of dust, above $10^7M_{\odot}$ \citep{Ber03,Car04,Fan06,Wi10}. 
Recently, \citet{Za15} measured for the first time a dust mass of $10^7M_{\odot}$ at redshift $\sim9$.

Despite of new observations and theoretical improvements, the origin of dust in such high redshift 
objects it is not yet understood:
this issue represents a very important point in understanding dust formation and chemical 
enrichment of the Universe \citep{Dw07,Ca14,Va14}. 
The fast dust production by Type II SNe may represent the primary channel of dust source, 
but also AGB or super-AGB stars have a non-negligible role \citep{Va09}.
The role of dust accretion is also uncertain: 
this process seems fundamental to match observations \citep{Mancini2016,Gin17},
but on the other side, it encounters several problems such as a too high temperature or repulsive Coulomb barriers that may reduce or stop grain growth \citep{Fe16}.
A top heavy IMF has also been proposed to easily solve the ''dust crisis'': 
in this scenario, a larger number of massive stars can sensibly increase the dust mass in the ISM \citep{Ro14}.

The problem of the origin of dust at high redshift becomes even more complex 
if one intends to study its evolution until the present time.
Recently, \citet{Po16} used a semi-analytical model of galaxy formation 
to study the dust content in cosmological volumes: 
their study confirmed that dust accretion is a necessary ingredient to reproduce the buildup of dust in galaxies.
\citet{McKinnon16} studied the distribution of dust in a cosmological hydrodynamical simulation: 
their model fails in reproducing dust-rich galaxies at high redshift, even when a top heavy IMF is adopted,
but they obtained a good agreement with the low redshift observations of the comoving dust density $\Omega_{dust}$. 

Several other multi-zone models have
included complex physical processes occurring in the 
different phases of the ISM \citep{Be13,Zhu16}, as well as the evolution of large and small grains \citep{Ao16}.
Beside hydrodynamical simulations, chemical evolution models are very useful to understand the behavior of dust in the ISM 
of single galaxies \citep{Dw98,Ca08,Zhu14,Bek14}.
By means of such models, \citet{Grieco14} used chemical evolution models with dust to compute
the cosmic dust rate (CDR) 
as the result of the contribution of galaxies of different morphology.
For this purpose, they assumed two different scenarios of galaxy formation,
which mimic the monolithic and hierarchical 
formation theories: the pure luminosity evolution scenario 
(were the number density of galaxies is constant) and the number density evolution scenario, respectively.  
Their estimate of the dust comoving density was larger than observed and the two different scenarios 
led to a very different behavior of the dust rate as a function of redshift.

In this work, we use the same method adopted by \citet{Grieco14} but with noticeable improvements:
first, we constrain our reference model of irregular, spiral and elliptical galaxies by the comparison with 
the observed dust amount in each galaxy type; second, we use a chemical evolution model with updated and detailed dust prescriptions, 
presented by \citet{Gi17} and recently adopted for the Milky Way in 
\citet{Spi17}; third, we explore more evolutionary scenarios of galaxy formation than in \citet{Grieco14}. 
Furthermore, since the dust content in galaxies is associated to star formation 
and to the presence of metals in the ISM, 
we compute also the cosmic metallicity evolution as well as the cosmic star formation rate (CSFR).

This paper is organized as follows: 
in Section~\ref{sec_chem} and~\ref{sec_dust_pres} we describe our new chemical evolution models with dust and 
the adopted dust prescriptions.
In Section~\ref{sec_Morph} we show the comparison between our models and observational data, 
and we define the reference models of a typical irregular, spiral and elliptical galaxy.
In Section~\ref{sec_cosmic_rates} we present the results of the evolution of the CDR, CSFR and the metallicity evolution. 
Finally, we draw our conclusions in Section~\ref{sec_conclusion}.

\section{Chemical evolution of galaxies}\label{sec_chem}
Galactic chemical evolution models are able to follow 
the evolution of gas chemical abundances during the life of a galaxy. 
In this paper we study the evolution of galaxies of different morphological type: 
ellipticals, spirals and irregulars. 
The set of the model parameters for different galaxies is constrained by previous studies 
which reproduce the chemical abundance patterns, observed average metallicity and constraints 
such as star formation rate, SN rates and present day gas mass.
We assume the galaxies to form by infall of primordial gas in a pre-existing diffuse dark matter halo, 
with a mass of about 10 times the total mass of the galaxy. 
In all the models, the stellar lifetimes are taken into account, thus relaxing the instantaneous recycling approximation.

In the case of elliptical galaxies, we consider a short time-scale of the infall ($\tau_{inf}$) 
as already adopted by previous works \citep{Ma_To87,Ma94}.
This assumption causes a rapid collapse, which triggers an intense and rapid star formation process.
As the gas thermal energy, heated by SN explosions, equates the binding energy of the system, 
an intensive galactic wind is produced and it stops the star formation and sweeps out all the residual ISM. 
Following this moment, the galaxy is assumed to evolve passively. 

Models of spiral galaxies are usually focused on reproducing observational constraints of 
the Milky Way. 
\cite{Ch97} described the formation of our Galaxy as a result of two main infalls which gave rise first to the galactic halo and thick disk (1 Gyr)
and later to the thin disk (8 Gyr).
More recently, \cite{Mi13} computed a three-infall model, where the formation of the halo, the thick disk and the thin disk
are originated in three different gas accretion episodes with time-scales of $\sim0.2, \sim1.25$ and $\sim6$ Gyr, respectively.
The observational data we used usually describe disk-like galaxy, so here we adopt for spirals a one infall model with 
the typical timescale for the formation of the thin disk.

Concerning irregular galaxies, they are assumed to assemble with a long and continuous star formation and a 
large infalling timescales \citep{Br98,Re_02,Gre04}.
In this way, they have a moderate star formation history which leads to a slower evolution with respect to the previous two types.

In general, our prescriptions lead to a galactic evolution in which more massive galaxies have higher star formation.
In particular, the star formation rate at the present time represents one of the observables that the model reproduces 
(as we will show later in the top right panel of Fig.~\ref{fig:Rates}).
Another important parameter that in our work differs between galaxies of different morphological type is the adopted initial mass function (IMF): 
we will explain the reason of our choice in the next paragraph.

Finally, the selected parameters used in this work also explain the mass-metallicity relation for galaxies of different morphological type, 
for which a detailed description can be found in \citet{Ca09}.

\subsection{Equations}
In this paragraph, we briefly describe the chemical evolution models used in this work. 
We refer the reader to \citet{Gi17} for a more detailed description of the model.

Let us define  $G_i(t)=M_{ISM}(t)/M_{ISM}(t_G)$ as the mass fraction of an element $i$
in the ISM at the present time, over the total gas mass.
The temporal evolution of $G_i(t)$ is determined by the following equation:
\begin{equation} \label{eq_chem_mod}
\dot {G}_{i}(t) = -\dot G_{i}^{SFR} + \dot G_i^{production} + \dot {G}_{i}^{infall} - \dot {G}_{i}^{wind}
\end{equation}
$\dot G_{i}^{SFR}=-\psi(t)X_i(t)$ represents the fraction of an element $i$ removed from the gas by star formation, 
where $\psi(t)$ is the star formation rate and $X_i(t)=G_i(t)/G(t)$ is the abundance by mass. 
The expression we assume for the SFR is a characteristic Schmidt law \citep{Sc59}: 
\begin{equation}
 \psi(t)=\nu G(t),
\end{equation} 
where $\nu$, in units of $Gyr^{-1}$, is the star formation efficiency and expresses the rate at which the stars form.

The second term of Eq.(\ref{eq_chem_mod}) $G_i^{production}$ represents the rate at which the mass fraction of the element $i$ is restored into the ISM by stars. 
This term takes into account the chemical enrichment of single low and intermediate-mass stars ($0.8<m_*/M_{\odot}<8$), 
core collapse SN explosions of massive stars ($8<m_*/M_{\odot}<80$), and 
Type Ia SNe, for which we assume the single-degenerate (SD) scenario. 
In this scenario, a C-O white dwarf in a binary systems accretes mass from the less massive and 
non-degenerate companion, until a mass of $\sim1.4M_{\odot}$ is reached, causing its explosion via 
C-deflagration \citep{Wh_Ib_73,Ma_Re01}. 
The mass and the chemical composition injected into the ISM by a stellar population depends on 
the initial mass function (IMF)
and on the adopted the stellar yields: the latter quantities represent the amount of both newly 
formed and
pre-existing elements injected into the ISM by stars at their death. In this work, we adopt 
the same stellar yields as adopted in \citet{Gi17}.
The third term on the right side of Eq. (\ref{eq_chem_mod}) accounts for the infalling material 
which accretes onto the galaxy.
As we already said, we consider an exponential law for the infall, where $\tau_{inf}$ is the characteristic time-scale:
\begin{equation}
 \dot G_{i}^{infall} = X_{i,infall}exp^{-t/\tau_{inf}},
\end{equation}
where $X_{i,infall}$ describes the chemical abundance of the element $i$ of the infalling gas, assumed to be primordial.
The last term in Eq.~\ref{eq_chem_mod} represents the element mass fraction removed from the ISM by the galactic wind.
This term is proportional to the SFR through the dimensionless parameter $\omega_i$, which is the efficiency of the wind for a specific element i:
\begin{equation}
 \dot {G}_{i}^{wind}=\omega_i \psi(t).
\end{equation}
Another important quantity of the model is the stellar initial mass function (IMF): 
it represents the mass distribution of stars at birth in a stellar population. 
In this work, the IMF is assumed to be constant in space and time and normalized to unity in the total mass interval considered.
In particular we adopt a single slope IMF \citep{Sa95} for elliptical and irregular galaxies:
\begin{equation}
 \phi_{Salp} (m)= 0.17 \cdot m^{-(1+1.35)}
 \end{equation}
This choice better reproduces the color-magnitude diagram of elliptical galaxies \citep{Pi_Ma04} and the metal content in cluster of galaxies \citep{Ca_Ma_To07}.
Furthermore, \citet{Ca04} showed that the application of a steeper IMF in dwarf irregulars under-estimates their average metallicities. 
On the other hand, for disc-like galaxies, we adopt a two slope approximation of the \cite{Sc86} IMF:
\begin{equation}
\phi_{Scalo}(m)=
\begin{cases} 0.19 \cdot m^{-(1+1.35)}, &  for \,\,\,\, m<2 M_{\odot}  \\ \\
0.24 \cdot m^{-(1+1.70)}, &  for \,\,\,\, m>2 M_{\odot}  ,
\end{cases}
\end{equation}
which is preferred with respect to the Salpeter one for spiral discs \citep{Ch01}. 
A detailed discussion on this topic in the framework of chemical evolution models can be found in \citet{Ro_05}.

\section{Dust prescriptions}\label{sec_dust_pres}
In this Section, we present the dust prescriptions adopted for the different galaxy models. 
In particular, we remind here some important features of dust modeling which are important for 
this paper. 
A more detailed explanation of the model can be found in \citet{Gi17}.

The AGB stars and Type II SNe play a fundamental role in the so called ''dust cycle'',
by forming dust which is injected in the ISM. 
We define the evolution of the dust rate of an element $i$ as below: 
\begin{equation}\label{dust_model}
 \dot {G}_{i}(t) =  \dot G_{i,dust}^{production} + \dot {G}_{i,dust}^{accretion} -\dot G_{i,dust}^{SFR} \\ 
 -\dot {G}_{i,dust}^{destruction} - \dot {G}_{i,dust}^{wind}
 \end{equation}
This equation includes all the processes occurring in the dust cycle. 
From the first term to the fourth on the right hand side of equation~\ref{dust_model}, they 
describe dust production by stars,
dust accretion in the ISM, dust astration, which accounts for the removal of dust from the ISM 
to form new stars and
dust destruction, respectively. Finally, the last term of Eq.~\ref{dust_model} accounts for the 
mass of dust lost by means of the galactic wind. 

\subsection{Dust Production}\label{sec_d_prod}
It has been recognized that mass-loss and dust formation occur during the late evolutionary phases of AGB stars \citep{Ge89}. 
The cold envelope of AGB stars, accompanied by its thermal pulsations,
provides a good environment in which dust nucleation can occur \citep{Ga09}. 
The chemical composition of the dust in the circumstellar shells of these stars 
is deeply affected by the stellar composition. 
Therefore, the better we know AGB stars, the better constraints can be inferred on dust composition \citep{Na13,Ve12,Ve14}. 
In particular, a very important feature is the C/O ratio present in the stellar surface. 
Depending on this ratio a carbon-rich star (C/O>1) produces carbon-enhanced dust whereas a 
oxygen-rich star (C/O<1) or a S-star (C/O$\sim$1) preferably produces other compounds, which we include in silicates. 
\cite{Fe_Ga_06} provided dust yields on the basis of the characteristics of AGB stars (mass and metallicity). 
In this article, we adopt condensation efficiencies provided by \cite{Pi11}, adopted already in \cite{Gi17}, which depend on the mass and the metallicity of the star.

Condensation efficiencies represent the fraction of a single element expelled by a star of a given mass and metallicity which goes into the dust phase.
By adopting these prescriptions in our models we find a good agreement with other studies of chemical evolution,
such as \citet{Dw98,Zh08,Va09}: 
low mass stars ($\le 3M_{\odot}$) mostly produce carbon dust, while silicates become important 
in the high metallicity range or when the mass of the progenitor is higher (up to $8M_{\odot}$).

Type II SNe are also important dust sources beside AGB stars: 
the total dust mass injected into the ISM from these kind of stars is larger than in AGB stars, 
due to the higher masses involved in the process. 
The direct evidence began from the observation of SN1987A \citep{Lu89,Da91}. 
SN1987A represents the best observable we have for such a kind of SNe:
recent observations in the mm and sub-mm bands \citep[PACS/SPIRE, Herschel Space Observatory]{Ma11,Ma15} \citep[ALMA]{In14}) 
revealed the presence of a cold dust component which increased the previous measurements up to a 
total mass of dust of about $0.7M_{\odot}$. 
In spite of this, 
the real composition and the amount of dust originating from these SNe is still not understood. 

In our work we adopt condensation efficiencies for Type II SNe as provided by \cite{Pi11}. 
The condensation efficiencies $\delta_{SNII}$ depend on the density of the environment surrounding the explosion:
in general, the higher the density, the more resistance the shock will encounter, 
and more dust will be destroyed by the reverse shock of the SN.  
In \cite{Gi17}, we adopted a median value of $n_H=1~cm^{-3}$, while
in this work we will explore the consequences of adopting different values for this quantity.
We will call 
the highest condensation efficiency value  
$\delta_{HP}^{SNII}$ and the lowest value $\delta_{LP}^{SNII}$, while the middle value is $\delta_{MP}^{SNII}$.
The effects of varying the condensation efficiency will not be studied in spiral galaxies since, as discussed in \cite{Ca08},
their dust depletion pattern is mostly regulated by the balance between accretion and destruction and generally insensitive
to the prescriptions regarding stellar dust production. This is particularly true after the critical metallicity has
been reached (see Sect.~\ref{sec_ref}).

To summarize, we can define the dust production from stars as the sum of the contribution from AGB and Type II SNe as:
\begin{equation} 
 \dot G_{i,dust}^{production} = \dot G_{i,dust}^{AGB}(\delta_AGB) + \dot G_{i,dust}^{SNII}(\delta_SNII), 
\end{equation}

where $\delta^{AGB}$ and $\delta^{SNII}$ represent the dust condensation efficiencies for AGB and Type II SNe, 
respectively.
We exclude the contribution from Type Ia SNe, 
as there is no firm observational evidence that they may produce a significant amount of dust (see \citealt{Gi17}).
 
\subsection{Dust accretion and destruction}
Dust accretion, or grain growth, may occur in the coldest and densest regions of the ISM,
such as molecular clouds. This mechanism increases the dust mass in a galaxy 
besides production by stars.
The analytical formula for dust accretion $G_{i,dust}^{accretion}=M_{dust}/\tau_{acc}$ comes from previous studies
of \citet{Hi00,As13}, as explained in \citet{Gi17}. 
Assuming 50 K as the reference value for the temperature of clouds, a hydrogen ambient 
density of $n_H=1.0~cm^{-3}$ and an average value of $0.1~\mu m$ for grain size,
we obtain for the accretion time-scale:
 \begin{equation}\label{Asano_accr}
 \tau_{acc,i}= 2.0 \times 10^7 yr 
 \times \dfrac{1}{X_{cl} (1-f_i)}
 \times \left(\dfrac{Z}{0.02}\right)^{-1} 
\end{equation}
Where $X_{cl}$ represents the mass fraction of cold clouds in a galaxy and $f_i=G_{dust,i}/G_i$
is the dust-to-gas ratio for the $i-$th element.

The other important process which affects dust mass in the ISM is dust destruction. 
Several processes are responsible for destroying dust grains such as thermal sublimation, 
grain-grain collision (shattering) or thermal sputtering. 
However, the most efficient process able to cycle dust back into the gas phase is 
dust destruction in SN shocks, whose rate can be expressed as $G_{i,dust}^{destruction}=M_{dust}/\tau_{des}$. 
The time-scale for dust destruction is:
\begin{equation} \label{time-scale-destruction}
\tau_{des}=\dfrac{M_{ISM}}{(\epsilon \cdot M_{Swept})SN_{rate}}= \dfrac{M_{ISM}}{1360\cdot SN_{rate}}
\end{equation}
where $M_{Swept}$ is the mass of the ISM swept out by the SN remnant which,
according to \citet{As13}, is metal-dependent and can be described by the following expression:
\begin{equation}\label{sweptup}
 M_{Swept}=1535\cdot n_H^{-0.202}\cdot[Z/Z_{\odot}+0.039]^{-0.289} [M_{\odot}].
\end{equation}

\section{Results: dust formation and evolution in galaxies}\label{sec_Morph}
In this Section we present the comparison 
between the results of our models 
and observational data related to the dust content of galaxies.
The dust-to-gas ratio versus metallicity is studied here in local irregular and spiral galaxies only.  
The large spread of \textrm{$\sim2~dex$} observed in the data can be interpreted as
the result of the peculiar star formation history of each single galaxy \citep{Re14},
as well as a different evolution of the dust \citep{Po16}.
Here we will show the effects on the dust-to-gas ratio by varying both the 
chemical evolution parameters and dust prescriptions.
 
Local elliptical galaxies generally show a modest content of gas and dust.  
Dust evolution in such systems will be discussed in Sect. \ref{sub_ell},
where the dust content of high redshift galaxies will be studied.

In Table~\ref{tab:grid} we show the range of the parameters we considered for each galaxy type.
In the first column the infall mass $M_{infall}$ is reported (this is the assumed mass of gas which assembles to form a given galaxy),
in the second the infall timescale $T_{infall}$, 
the star formation efficiency $\nu$ in the third column, 
the wind efficiency $\omega_i$ in the fourth, in the fifth the efficiency of dust production by Type II SNe
and in the sixth our prescriptions regarding dust accretion, i. e. whether it is taken into account or not. 
We emphasize that we varied the parameters in the range allowed by previous works of chemical evolution models,
able to reproduce the average stellar abundances and the mass-metallicity relation, as explained at the beginning of Section \ref{sec_chem}. 
These parameters are not tuned to reproduce the observed dust properties. 
The aim of the present work is to study in which way these parameters infuence the behavior of the dust mass in the ISM.

 \begin{table*}
 \begin{center}
  \renewcommand\arraystretch{1.5}
  \begin{tabular}{c c c c c c c}
  \hline
   Type & $M_{infall}[M_{\odot}]$ & $T_{infall}[Gyr]$ & $\nu[Gyr^{-1}]$ & $\omega_{i}$ &  $\delta_i^{SNII}$	& Accretion\\
   \hline
   Irregular & $10^9\leq M\leq10^{10}$  & $0.5 \leq T_{infall}\leq10$ & $0.01\leq \nu\leq0.2$	& $0.1\leq\omega_i\leq1.5$ & $\delta_{LP,\,\,MP,\,\,HP}^{SNII}$ & yes \\
   Spiral    & $5\times 10^{10}$      & $7.0$	    & $1.0\leq \nu\leq3.0$	& $0.1\leq \omega_i\leq1.0$ & $\delta_{MP}^{SNII}$& yes	\\
   Elliptical& $10^{11}\leq M\leq10^{12}$  & $0.2\leq T_{infall}\leq1.0$	    & $10\leq \nu \leq20$		& $10$ 		     & $\delta_{LP,\,\,MP,\,\,HP}^{SNII}$ & yes-no	\\
  \hline
   \end{tabular}
   \label{tab_variation}
\caption{Parameter ranges used in our work for the chemical evolution models of irregular, spiral and elliptical galaxies. 
In the first column we report the morphological type of the galaxy,
in the second, third, fourth and fifth column the $M_{infall}$, $T_{infall}$, 
$\nu$ and $\omega_{i}$, respectively.
In the sixth and seventh columns the choices for the Type II SNe condensation efficiencies and dust accretion are shown, 
respectively. }\label{tab:grid} 
 \end{center}
\end{table*}

 \subsection{Irregular and spiral galaxies: the dust-to-gas ratio in the Local Universe}\label{sub_spi_irr}

 \begin{figure*}
\centering
 \includegraphics[width=0.49\textwidth]{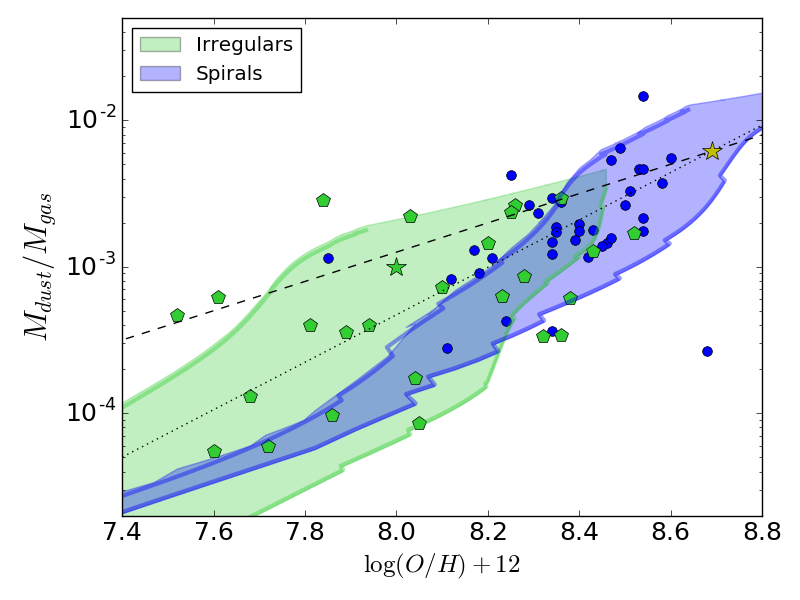}
 \includegraphics[width=0.49\textwidth]{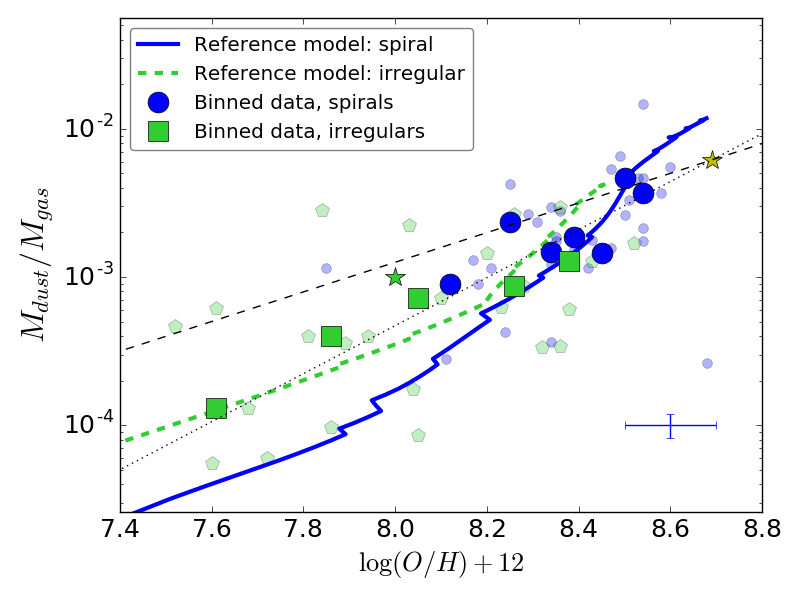} 
\caption{Dust-to-gas ratio versus $\log(O/H)+12$. 
Data:
Green pentagons correspond to irregular galaxies, whereas blue filled circles correspond to spirals. 
Yellow and green stars represent the dust-to-gas ratio observed in the solar neighborhood  
 \citep{Zu04} and in the Small Magellanic Cloud \citep{Le07}, respectively. 
The dashed-black line expresses a linear relation between dust-to-gas and metallity, whereas the dotted straight line 
 is the fit of the whole data sample as done in \citet{Re14}.
Left panel: green and blue areas represent the possible values of dust-to-gas ratio and metallicity 
achievable by our models by varying the model parameters of Table~\ref{tab:grid}, for spirals and irregulars. 
Right panel: predictions by the reference models of a typical irregular (dashed green) and spiral (solid blue). 
Here, data for irregulars and spirals have been binned in metallicity, with a minimum dataset of 5 points in each bin
(large green squares and blue circles, respectively). Typical errors in dust-to-gas and metallicity are expressed by the blue cross
in the bottom-right corner. 
}\label{fig:DtG}
 \end{figure*}

 \subsubsection{Data sample}
To study the observed spread in the dust-to-gas ratio, we used in our work  
a subsample of the Dwarf Galaxy Survey \cite[DGS]{Ma13} and the KINGFISH survey \citep{Ke11}. 
The information about the gas and dust content in these galaxies can be found in \citet{Re14,Re15}. 
The DGS sample mostly consists of dwarf galaxies in the local Universe, also 
including irregulars and blue compact dwarfs. 
The main physical parameters of these galaxies can be found in \citet{Ma13}: the sample covers the low metallicity range,
from $12+\log(O/H)=7.52$ up to 8.43. 
On the other hand, the KINGFISH sample contains more metal-rich galaxies, which are spirals.
Contrary to previous studies, we build two subsamples, according to the different morphological type of these galaxies.

In the plots we are going to present we will refer to the metallicity of the gas expressed as $\log(O/H)+12$,
as explained in ~\citet{Re14}, in both data and models.
The dust masses provided by \citet{Re15} are constrained on the basis of the spectral energy distribution of each galaxy.
They provide two different values for the dust mass depending on the assumed dust species forming the carbonaceous grains:
when carbon dust is assumed in form of amorphous carbonaceous grains, the dust mass is about 2.5
times lower than the one obtained by a graphite-grain model.
Therefore, using different models
for carbonaceous grains modifies the absolute values
of dust masses only, leaving unchanged the observed trend in the
dust-to-gas ratio versus metallicity. 
As we are mostly interested in this trend, we compare our models with the average values of the two measurements.

\subsubsection{Chemical evolution models as constrained by data}
We have built a grid of models by varying the main parameters in order to study the observed spread of 
the dust-to-gas ratios in irregular and spiral galaxies (Table~\ref{tab:grid}). 
The variation of some of these parameters affects significantly the computed dust-to-gas ratio and
the metallicity. 

In the left panel of Fig.~\ref{fig:DtG} we show the model results and the observational data in the gas-to-dust ratio versus 
metallicity space:
the green and blue areas represent the region occupied by models for various values of the parameters 
for the irregulars and spirals, respectively.
Irregulars mostly lie in a low metallicity range, with a median value of $[\log(O/H)+12]_{irr}=8.03$ 
and show a large spread of the dust-to-gas ratio. 
This is mainly due to the efficiency of dust production from Type II SNe: 
clearly, adopting different dust production by these sources 
leads to broad differences in the dust-to-gas ratio at low metallicities.  
On the other hand, the spread above $\log(O/H)+12\simeq8.0$ is not due to our dust prescriptions, but it can be attributed
to the variation of the time-scale of the infall or infall mass or galactic wind.

Spiral galaxies are more concentrated at higher metallicities, with a median value of $[\log(O/H)+12]_{spi}=8.37$.
Both models and data show a narrower dispersion of the dust-to-gas ratio with respect to irregulars.
In this latter case, the star formation efficiency is the main responsible for the observed spread above $\log(O/H)+12=8.4$.
In this picture, galaxies presenting similar values for the dust-to-gas ratio but different metallicities can be interpreted
as the result of the variation of their star formation histories.

In the right panel of Fig.~\ref{fig:DtG} we show the models we choose as representative cases of each morphological type, 
hence our reference ones.
In this case, we bin the data in metallicity, 
obtaining a good agreement between model results and data for both irregular and spiral galaxies.

In Table~\ref{tab_best} we report the main parameters characterizing our reference models, namely the mass, the infall time-scale, 
the star formation efficiency, the wind efficiency, 
the adopted IMF and the adopted dust contribution from Type II SNe.

\subsection{Elliptical galaxies: dust evolution in the high redshift universe}\label{sub_ell}
\begin{figure}
\centering
 \includegraphics[width=0.49\textwidth]{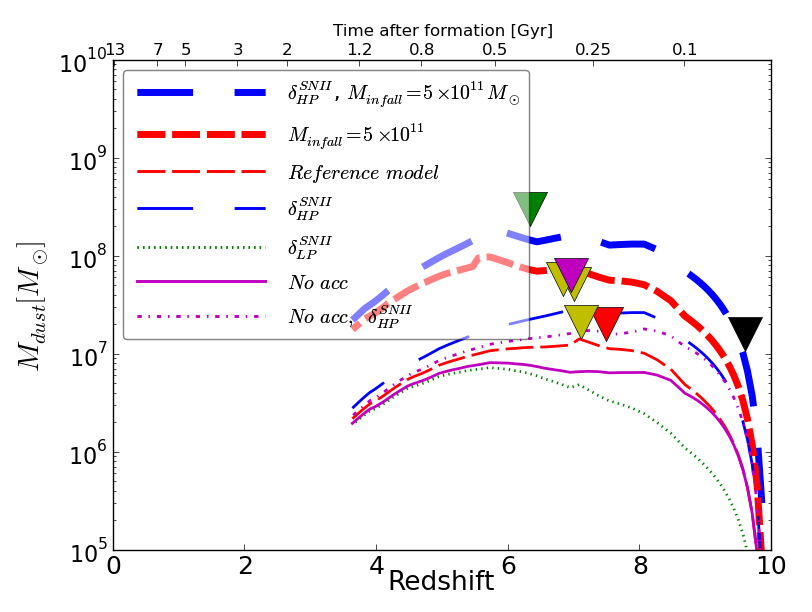}
 \caption{Dust mass in elliptical galaxies as a function of redshift (lower x-axis) and time (upper x-axis).
 The thin red long-dashed line is our reference model, whose parameters are shown in Table~\ref{tab_best}. 
 In the models described by the blue long-dashed and green dotted lines, higher ($\delta_{HP}^{SNII}$) and
 lower ($\delta_{LP}^{SNII}$) dust production by Type II SNe are adopted,  respectively. 
 The magenta solid and dot-dashed lines represent models with no accretion: 
 the former assumes a middle contribution from Type II SNe ($\delta_{MP}^{SNII}$), whereas 
 the latter a higher one ($\delta_{HP}^{SNII}$).
 The thick lines represent models with higher infall mass ($M_{infall}=5\times11M_{\odot}$).
The data represent upper limits for the 
 dust mass as observed in high redshift galaxies by \citet{Wa15} (red), \citet{Co14} (green),
 \citet{Za15} (black), \citet{Maiolino15} (yellow) and \citet{Ota14} (magenta).}\label{fig:ellipticals}
 \end{figure}

In this Section we will use our models for elliptical galaxies to investigate the origin of dust in high redshift star-forming galaxies, which is still a rather debated issue. 
Type II SNe and AGB stars \citep{Va09,Laporte17} have been regarded as possible dust sources, 
however, non-stellar sources seem to be required (e. g., \citealt{Matt11,Kuo12}),
as the contribution from stellar populations can hardly account for the amount of dust observed in some high redshift systems (e.g., \citealt{Pip11,Ca14}).
Such a ''dust-budget crisis''  can be solved by assuming a top-heavy IMF \citep{Ro14} or alternative channels for dust production. 
In \citet{Pip11} an ''extra QSO-dust'' source was introduced to explain the high amount of dust observed in QSOs.
\citet{Mancini2016} claimed that the dominant contribution to the dust mass, 
for galaxies above z>6 comes from grain growth, but  
on the other hand, this process might be delayed because of the high temperature in the high redshift ISM \citep{Fe16}.

With our model we are able to differentiate the contributions of 
different channels of dust production, and therefore study the mechanisms of dust formation in high redshift objects.

In Fig.~\ref{fig:ellipticals} we show the comparison between our models and dust mass measurements in high redshift galaxies, 
by assuming, hereafter, a $\Lambda$-cold dark matter ($\Lambda$CDM) cosmology with $H_0=67.7 Km\,s^{-1}\,Mpc^{-1}$ \citep{Pl16} and redshift of galaxy formation z=10.
The reference model (red-dashed line) lies below the upper limits fixed by observations and it reaches a total mass of dust  
$M_{dust}\ge10^7M_{\odot}$ at its maximum. 
The model with a higher dust production from Type II SNe ($\delta_{HP}^{SNII}$, thin blue dashed)
leads to a larger amount of dust, starting from the earliest phases of its evolution. 
In order to reproduce the measured dust masses, our model needs the most efficient contribution
from Type II SNe as well as a higher mass of the infall ($M_{infall}=5\times10^{11}M_{\odot}$, thick blue long-dashed line): 
in this scenario a significant dust mass is produced even at $z\simeq9.6$ \citep{Za15} and reaches values $M_{dust}>10^8M_{\odot}$ 
for $z\le6$ \citep{Co14}.
A similar result is obtained by the red thick dashed line, which represents the same prescriptions as the reference model,
but with $M_{infall}=5\times10^{11}M_{\odot}$.

Finally, we also tested the case with no dust accretion (magenta dash-dotted line):
in this case, it is impossible to reproduce high values for $M_{dust}$ at $z>6$.
If we consider the highest contribution by Type II SNe (magenta dash-dotted line), 
in the earliest phases the dust mass budget is even larger than the one predicted by 
the reference model.

Our work also casts light on another important issue related to high redshift galaxies:
the time-scale of dust formation. 
The simultaneous presence of both dust-rich and dust- (and metal-) poor objects (e.g. \citealt{Ou13}) at high redshift 
indicates that the transition between these two populations
has to be very rapid and may take place on a Myr time-scale \citep{Matt15}.
Fig.~\ref{fig:ellipticals} is very useful to understand which is the timescale of the the buildup of
dust, as well as by which process it is driven.
Fig.~\ref{fig:ellipticals} shows that in the earliest phases, i.e. at times $\le ~0.1$ Gyr, the fast increase of the dust mass
is mainly driven by the contribution of Type II SNe. This is visible from, e.g., the lack of any
difference in the behavior of the reference model and the 'no accretion' model
at these times. 

Other works have already shown that in elliptical galaxies, the buildup of the metals occurs
on a rapid timescale, with a supersolar metallicity reached already at  $\sim~0.1$ Gyr (e.g., \cite{Ca14}). 
As heavy elements are the main constituents of dust grains, dust accretion can start to be significant 
only after enough metals have been ejected into the ISM and are available for coagulation
onto pre-existing grains, and this occurs after $\sim~0.1$ Gyr. 
Moreover, the bulk of dust mass is settled already at $\sim~0.25$ Gyr after the beginning of the star formation. 
Clearly, in proto-spheroids the rapid buildup of the dust is also helped by the rapid time-scale of the infall. 
The bulk of the dust is forming during the burst of star formation, where Type II SNe 
have the most important contribution. 
In fact, in the quiescent phases, AGB stars produce dust in a negligible fraction with respect the one formed during the initial burst, 
even if the contribution of Type II SNe and AGBs strongly depends on the adopted yields. 
This result is different than previous works where a non-negligible mass of dust was assumed to form
after the onset of the galactic wind due to the contribution of Type Ia SNe \citep{Ca08,Ca17}, 	
which in this model are not considered as dust factories.

 \begin{table*}
 \begin{center}
  \renewcommand\arraystretch{1.5}
  \begin{tabular}{c c c c c c c}
  \hline
   Type & $M_{infall}[M_{\odot}]$ & $T_{infall}[Gyr]$ & $\nu[Gyr^{-1}]$ & $\omega_{i}$ & IMF & $\delta^{SNII}$		\\
   Irregular & $10^{10}$  		 & $5.0$ 	    & $0.2$		  & $0.7$ 	  & Salpeter   	& $\delta^{SNII}_{HP}$ \\
   Spiral    & $5\times10^{10}$		 & $7.0$	    & $2.0$		  & $0.3$ 	  & Scalo	& $\delta^{SNII}_{MP}$		\\
   Elliptical& $10^{11}$  		 & $0.3$	    & $15$		  & $10$ 	  & Salpeter	& $\delta^{SNII}_{MP}$		\\
  \hline
   \end{tabular}
\caption{Parameters of the reference models for galaxies of different morphological type.
In the first column the morphological Type is shown, in the second one the mass of the infall (in solar masses),
in the third the infalling time-scale (in Gyr), in the fourth the star formation efficiency (in $Gyr^{-1}$),
in the fifth the wind parameter, in the sixth the adopted IMF and in the seventh the efficiency 
of dust production by Type II SNe.}\label{tab_best}
 \end{center}
\end{table*}

\begin{figure*}
    \centering
      \includegraphics[width=0.78\textwidth]{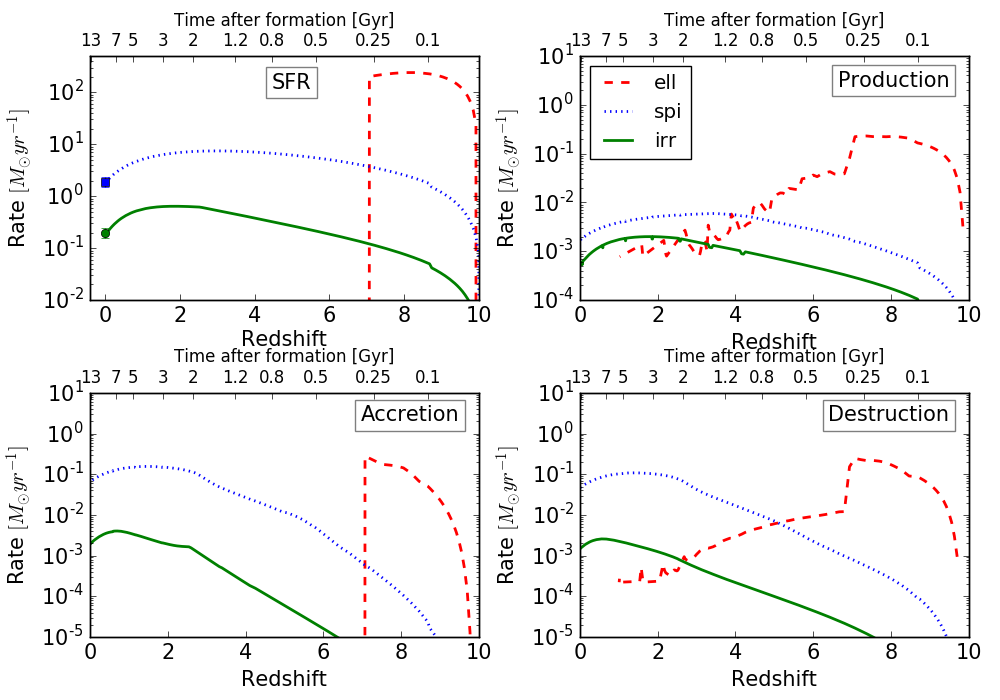}
\caption{In the top left panel, 
the red dashed, blue dotted and green solid lines show 
the star formation rate as a function of time and redshift for our chemical evolution models of an elliptical, a spiral and an irregular galaxy, respectively. 
All galaxies are assumed to form at redshift 10.
Present-time data on the SFR are taken from \citet{Ch_Po_11} (blue square) and \citet{Ha_Za_09} (green circle) 
for our Galaxy and the Large Magellanic Cloud, respectively.
In clockwise sense, the other panels of the figure show the evolution of the dust production, destruction and accretion.}
\label{fig:Rates}
\end{figure*}

\subsection{The reference models}
\label{sec_ref}

We present in Fig.~\ref{fig:Rates} the redshift evolution of an irregular, spiral and elliptical galaxy 
which we assume to form at redshift 10. 
The top left panel shows the SFR: the elliptical galaxy consists in an intense burst of SF, which stops 
before 1 Gyr of evolution due to the occurrence of galactic winds. 
The SFR in the initial phase is higher than the one of the irregular and spiral, according to the downsizing scenario 
(i.e. higher star formation efficiency in larger galaxies, \citealt{Ma94}). 
The spiral galaxy is characterized by a continuous SFR which is higher than the one of irregulars.
The models are able to reproduce the present day SFR observed in irregulars \citep{Ch_Po_11} and spirals \citep{Ha_Za_09}.
In the other panels of the Figure,
the evolution of various processes related to dust evolution (stellar production, accretion, and destruction) are shown.

The stellar dust production rate (top-right panel of Fig.~\ref{fig:Rates}) shows a trend which is very similar trend
to the SFR: 
this is because during the bursts, Type II SNe represent the dominant dust process
and the rate of this kind of SNe traces the SFR. 
The contribution of AGB stars is mostly visible in the quiescent phases, i. e. 
at the end of the burst of star formation of the elliptical galaxy.
In the lowest panels, the dust accretion and destruction rates are presented. 
In the cases of the spiral and irregular, dust production by stars has a dominant role at early epochs, 
while dust growth in the ISM becomes the most important process after the critical metallicity 
is reached\footnote{The Critical metallicity is the metallicity at which the contribution of dust 
accretion overcomes the dust production by stars \citep{As13}}.
In our models we found a critical metallicity of $Z_{crit,irr}=0.31Z_{\odot}$ and 
$Z_{crit,irr}=0.19Z_{\odot}$ for irregulars and spirals, respectively. 
Dust growth is strictly related to star formation and 
in the elliptical model it stops as soon as there is no more star formation.
In this case, at the end of the initial burst, 
the stellar production is still the most efficient process, and for this reason,  
the critical metallicity is not reached.
At variance with dust accretion, dust destruction never stops, 
but it has a marginal role in the case of irregulars and spirals.

 \section{Cosmic Rates}\label{sec_cosmic_rates}
In this Section we present our study on the evolution of the cosmic rates of star formation, dust production and cosmic mean metallicity. 
In general, a cosmic rate is defined as the rate in a comoving volume of the Universe:
it is the result of the contribution of galaxies of different morphological type, 
which should be weighted according to their number densities.
Since now on, we will assume that our models represent average galaxies belonging
to each morphological type, 
i.e. irregulars, spirals and ellipticals. 
We also tested the presence of bulges, which can be treated as elliptical galaxies but with lower masses  
than our reference model. 
The lower mass involved in this kind of objects leads to a negligible dust contribution
with respect to ellipticals, 
and no deviation from the presented results have been found.

Our method is based on the evolution of the galaxy number density,  
as already adopted in previous studies concerning 
cosmic star formation rate \citep{CM03,Vincoletto12}, cosmic SN rates \citep{Grieco12,Bon13}
and cosmic dust production  \citep{Grieco14}.
 
It is the rate at which galaxies have formed during cosmic time which determines 
the growth of baryonic structures in the Universe. 
The study of the cosmic rates can give important constraints on galaxy formation mechanisms, such as 
the monolithic collapse (MC) and hierarchical clustering (HC) scenarios.
In the first scenario, violent bursts of star formation originate spheroids and bulges at high redshift:
massive galaxies form with higher star formation efficiencies with respect to lower ones, 
reproducing the ''downsizing scenario'' \citep{Ma94}.
On the contrary, in the HC scenario, spheroids and bulges mainly form at late epochs as a result of mergers of galaxies such as spirals. 

If we define $n_k$ as the number density of galaxies of the k-th morphological type, 
its redshift evolution can be parametrized by the following equation:
\begin{equation}\label{eq_number_density}
n_k=n_{k,0}\cdot(1+z)^{\beta_k}
\end{equation}
where $n_{k,0}$ is the number density at z=0 and $\beta_k$ represents how the number density
(or the luminosity function,) evolves. 
The parameters we adopted are shown in Table \ref{tab_number_density}: these values have been taken from \citet{Vincoletto12}
which used a spectrophotometric code to reproduce galaxies of different morphological types
and were able to provide some constraints on the slope of the luminosity function.  
Thanks to Eq. (\ref{eq_number_density}), we can study the cosmic rates in two different extreme scenarios:
the ''pure luminosity evolution'' scenario (PLE), which consists in the case of no number density evolution ($\beta_k=0$),
and the ''density evolution'' one (DE), where the number density evolves with redshift ($\beta_k\ne0$).

We also present an \textit{alternative} scenario, based on the work 
of \cite{Po15} and able to reproduce the CSFR (see next paragraph).
In this scenario, the number density of spiral galaxies increases from z=0 up to z=2.3 
as in the DE model, but it decreases as an exponential law for higher redshifts: 
\begin{equation}\label{eq_alternative}
 n_{S}=n_{0}(1+z)e^{-(1+z)/2}
 \end{equation} 
In this context, at early cosmic times, the halo star formation phase is the dominant process in spiral galaxies,
whereas the disk star formation becomes dominant at lower redshifts.
Here, the disk formation is peaked at $z\sim2$, in agreement with recent cosmological simulations \citep{Murante15}.
Concerning ellipticals, their formation is assumed to start at redshift 5 and half of them form between the range $1\le z \le2$.

\begin{table}
 \begin{center}
  \begin{tabular}{c c c}
  \hline
   Galaxy Type & $n_0\, [10^{-3} Mpc^{-3}]$ & $\beta_k$ \\
   Irregular &   0.6	 &  0.0	 \\
   Spiral    & 	8.4	 &  0.9	\\
   Elliptical&   2.24	 & -2.5	 \\
  \hline
   \end{tabular}
\caption{Parameters adopted for describing the evolution of the number density 
for galaxies of different morphological type.}\label{tab_number_density}
 \end{center}
\end{table}

\subsection{CSFR: Cosmic star formation rate}\label{sec_CSFR}

\begin{figure}
    \centering
      \includegraphics[width=0.52\textwidth]{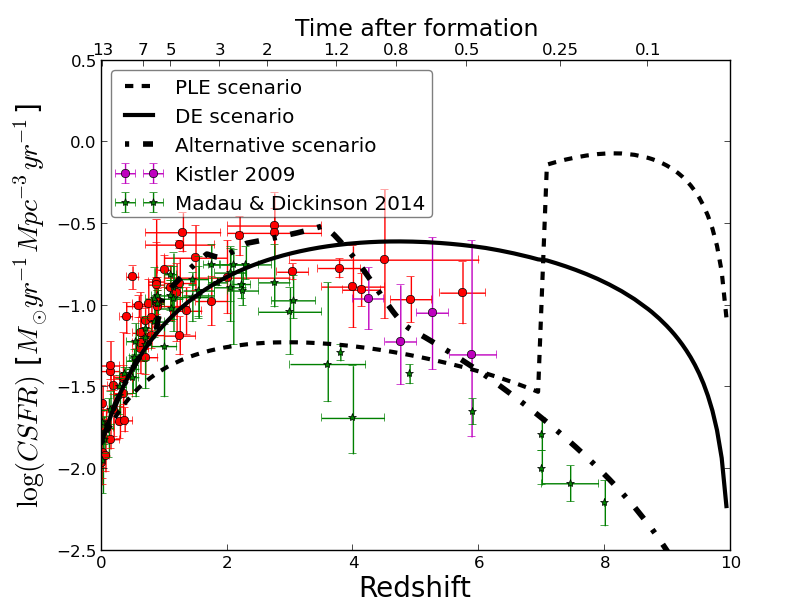}
       \caption{Cosmic star formation rate expressed in $[M_{\odot}yr^{-1}Mpc^{-3}]$ as a function of the redshift, 
       for PLE (black dashed), DE (black solid) and for the alternative (black dash-dotted) scenarios. 
       Green stars, red squares and magenta points data are taken from the compilation of 
       \citet{Ma_Di14}, \citet{Ho07} and \citet{Ki09}, respectively.}\label{fig:CSFR}
\end{figure}

The cosmic star formation rate (CSFR) is defined as the comoving space density 
of the global SFR in a unitary volume of the Universe. 
The CSFR can be defined as: 
\begin{equation}\label{eq_CSFR}
 CSFR=\sum_k \psi_k(t) \cdot n_k
\end{equation}
where $n_k$ and  $\psi_k(t)$ are the number density (as defined in eq. \ref{eq_number_density}) of galaxies and the
average star formation rate at the time $t$ for galaxies of the $k-$th morphological type, respectively. 
The above definition is necessary to study the redshift evolution in the
PLE or in the DE scenarios. 
The CSFR is manly provided by the integrated light of galaxies in the rest-frame
UV and IR wavebands~\citep{Ke98,Ke_Ev12}.
Measurements are affected by specific 
corrections related to the metal enrichment history or the choice of the IMF. 
In particular, they are deeply affected by the presence of dust, as it obscures 
UV region and produces IR emission~\citep{Ca01,Afo03}: this correction 
presents several uncertainties, especially when CSFR is only inferred from the UV light.
At low redshifts (0<z<1), the CSFR is inferred from IR measurements, but for
small comoving volumes over few independent sightlines. 
For higher redshifts, wider regions are covered, as the UV rest-frame light 
is visible using ground-based 
optical imaging (up to z$\simeq4$), but IR measurements are often unavailable~\citep[hereafter MD14]{Ma_Di14}.
For these reasons, CSFR measurements should be used with caution. 

Data from MD14 are shown in Fig.~\ref{fig:CSFR} together with the predictions of our models for the 
different scenarios: 
the data show a general increase of the CSFR until $z\simeq 2.5$ and then they start to decrease up to z=8.

In the PLE model (dashed line), the CSFR shows two peaks: a first, very broad one centered at $z\simeq2-3$,
and a much higher second one centered at $z\simeq9$. 
The latter peak is due to elliptical galaxies, 
which form with high star formation in a relatively short time-scale. 
As the burst of star formation ends, the CSFR decreases and spiral 
galaxies become the main drivers of the CSFR evolution until the present time. 
On the other hand, the DE scenario presents a smoother evolution: 
it is in good agreement with data at $z<3$, 
whereas it overestimates the observations at higher redshifts. 
In this case, the high redshift peak is absent because of the lower impact of elliptical galaxies. 
Despite of this, the CSFR still appears much higher than observations.\\
In Fig.~\ref{fig:CSFR}, we also show the CSFR as predicted in the alternative scenario in which the
formation of ellipticals occurs as suggested by \cite{Po15} (dash-dotted line). \\
\cite{Po15} used a 'backward' approach to 
interpret of the evolution of the near-IR and the far-IR luminosity functions (LFs) across the redshift range $0\le z\le 3$.
The spectral evolution of spheroids were described by a single-mass model, 
corresponding to a present-day elliptical with a K-band luminosity comparable to the one of the break of the local early-type luminosity function. 
\cite{Po15} 
used the redshift distributions and the source counts in the  near-IR and in the far-IR to constrain the main epoch of
spheroid formation, finding that roughly half of them must have formed in the redshift range $2\le z\le 5$. 

Assuming this scenario for the formation of elliptical galaxies, the CSFR at high redshift ($z>5$) is clearly dominated by spirals and
very well reproduced. 
At lower redshift, the combination between the formation of spirals and ellipticals leads the CSFR to peak at $z\simeq3$, and then to 
decrease until $z=0$. 

We suggest this as the best scenario. In fact, the DE scenario provides a good fit to the CSFR only at $z<2$ but
it predicts a too small density of ellipticals
($5.866\times10^5M_{\odot}Mpc^{-3}$) 
in the Local Universe with respect to the observational estimate ($1.84 \times 10^8 M_{\odot}Mpc^{-3}$, \citealt{Fu_Pe04}).
On the other hand, the alternative scenario provides a local stellar mass density in spheroids of $\sim 2 \times 10^8 M_{\odot}Mpc^{-3}$,
in agreement with the value calculated by \cite{Fu_Pe04}. 

As we have already pointed out, both observational and theoretical estimates of the CSFR are highly uncertain, because of the uncertainties in the dust extinction at high redshift. 
The large discrepancy between the observed CSFR at $z>6$ and
the one predicted in the PLE can be hardly explained 
by uncertain dust corrections only, but it is certainly possible that the observational values at high redshift could represent underestimates
of the real values. 

Some examples of CSFR measurements which differ substantially from the 
trend shown by the data collected by MD14 can be found in, e.g.,~\citet{Fau08} and 
~\citet{Ki09}. 

From their study of the evolution of the Lyman $\alpha$ effective optical depth, 
 \cite{Fau08} found a nearly constant 
intergalactic hydrogen photo-ionization rate
and cosmic star formation density across the redshift range $2\le~z~\le~4.2$.
Instead of a declining CSFR, these authors found a roughly constant value of $\sim 0.2 M_{\odot}~yr^{-1}~Mpc^{-3}$,
in contrast with the data plotted in Fig.~\ref{fig:CSFR}. 

Thanks to Gamma Ray Bursts (GRBs), ~\citet{Ki09} were able to constrain the CSFR at high redshift ($4<z<6$):
these measurements are sensibly higher than the data of MD14, and 
more in agreement with the DE scenario. 

In conclusion, in order to have a better agreement with the observed trend of the CSFR, 
we have to modify the number density evolution of spiral galaxies at redshift z$\ge$2.
However, due to the uncertainties in the inferred CSFRs, 
there is still the possibility of a high-redshift evolution different than that depicted by UV survey.
A clearer picture will hopefully be achievable in the future thanks to next generation telescopes, such as the
James Webb Space Telescope, which will allow one to detect weak H transitions insensitive to dust extinction up to very high redshift (MD14).

 \begin{figure}
    \centering
      \includegraphics[width=0.50\textwidth]{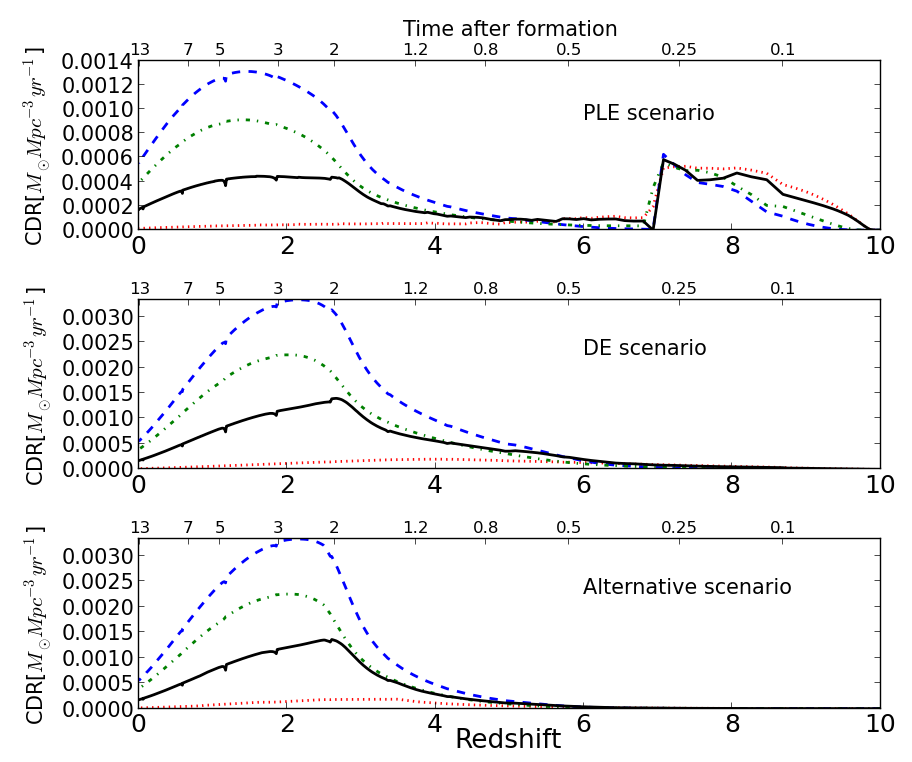}
       \caption{Cosmic dust rate versus redshift in the PLE (top panel), DE 
       (middle panel) and alternative (bottom panel) scenarios. The red dotted line refers to the dust production rate, 
       the blue dashed line is the  dust growth rate and the green dot-dashed line is the dust destruction rate.
       The solid black line represents the total dust rate. 
       }\label{fig:CDR}
\end{figure}

\subsection{Cosmic dust budget}\label{sec_CDR}

In this Section we present our results concerning the evolution of the global cosmic dust budget. 
From a theoretical perspective, an accurate assessment of the total amount of dust in the Universe is not an easy task,
owing to several uncertainties related to 
dust formation and destruction in different environments, 
such as the ISM, galactic halos, the intergalactic medium or high redshift QSOs \citep{Me_Fu12}. 
An attempt to assess the global amount of dust in the Universe on a phenomenological basis 
is the one by \citet{Fu11}, who took into account the dust present in galactic halos and disks. 

Observationally, an efficient method to derive constraints on the dust budget and its evolution is 
is by studying the cosmic far infrared background (CFIRB), powered by  
the UV and optical emission of young stars, absorbed by the dust in 
galaxies and then re-emitted in the infrared (IR) wavelengths, as done by 
\citet{De_Co12}. 

An accurate assessment of the evolution of the comoving dust mass density is the one performed by
\cite{Du11} by means  a wide galaxy sample from the Herschel-Astrophysical Terahertz Large Area Survey
(H-ATLAS; \citealt{Ea10}).
The sample, consisting of 1867 sources,  includes galaxies selected at 250 $\mu m$ presenting 
reliable counterparts in the Sloan Digital Sky Survey (SDSS) at $z < 0.5$. 
In the work by \cite{Du11} dust 
masses were calculated using both a single-temperature grey-body model for the spectral energy
distribution and a model with multiple temperature components, obtaining no significant differences
in terms of dust temperature evolution between the two methods. 
\cite{Du11} found a strong evolution of the dust mass density at redshifts $z \le 0.5$. 

Before discussing the global dust budget and its evolution, 
in analogy with  eq. (\ref{eq_CSFR}), we can define the cosmic dust rate (CDR) as:
\begin{equation}\label{eq_CDR}
 CDR=  \sum_k \lbrace \dot G_{i,dust}^{prod} + \dot {G}_{i,dust}^{acc} -\dot G_{i,dust}^{SFR} 
  -\dot {G}_{i,dust}^{des}\\
  - \dot {G}_{i,dust}^{wind} \rbrace \cdot n_k
  \end{equation}

In Fig.~{\ref{fig:CDR}, we show the evolution of the comoving cosmic production, accretion and 
destruction rates 
in the different scenarios introduced in Sect.~\ref{sec_cosmic_rates}. 
In the PLE case, we obtain two distinct peaks of dust formation, as already found in the evolution 
of the CSFR.  

In the DE scenario, the peak related to the initial bursts
of massive galaxies completely disappears as these objects form via merging of 
lower mass galaxies, for which the dust production rates are very low at high redshift. 
In this case, we obtain a  peak at redshift $2\le z\le3$ which is higher with respect to the one of the PLE scenario. 

The peak visible in the DE scenario reaches approximately the value 
obtained by the sum of the two peaks of the PLE scenario. 
For this reason, the total amount of dust produced at $z \sim 0$ (see Fig.~\ref{fig:Omega_dust})
is similar in the two cases, although their redshift evolution is totally different. 

On the bottom panel of Fig.~\ref{fig:CDR} we show the results for the alternative scenario. 
In this case, the evolution of the total dust rate is very similar to that obtained in DE model. 

The comoving dust mass density can be calculated as 
\begin{equation}\label{eq_CDD}
 \rho_{dust }=  \sum_k M_{dust,k} \cdot n_k, 
\end{equation}
where $M_{dust,k}$ is the average dust mass in galaxies of the $k-th$ morphological type at a given redshift. 
The dust mass density parameter  $\Omega_{dust}$ can then be derived by dividing $\rho_{dust}$
by the critical density of the Universe $\rho_{cri}=1.3\times10^{11}~M_{\odot}Mpc^{-3}$ \citet{Pl16}. 

In Fig.~\ref{fig:Omega_dust} we show the evolution of 
$\Omega_{dust}$ as a function of redshift as calculated by means of our models and compared to previous results from other works.
 
In both DE and alternative scenarios $\Omega_{dust}$ increases quite regularly from $z > 6$ to $z\sim 2$, where it peaks at a value of 
$\sim~4\times10^{-6}$, and then it starts decreasing to a final value of $\sim 7\times10^{-7}$. 
The only difference between the DE and the alternative scenarios concerns the evolution of $\Omega_{dust}$ at $z>3$, which is 
slightly steeper in the latter case. This difference reflects the behavior of the CSFR, 
which in the redshift interval $3 \le z \le 8$ appears to grow more steeply in the alternative scenario than in the DE one. 

In the PLE scenarios, the two peaks seen in the evolution of the cosmic dust rate
are reflected by the changes of slope in the $\Omega_{dust}$ vs $z$ plot.
The fast increase of $\Omega_{dust}$ at redshift $z>8$ reflects the formation of elliptical galaxies, which is followed
by a nearly constant evolution at $4\le z \le 8$, a second, considerable growth at $2\le z \le 4$ and a decrease at lower redshifts.  

All scenarios predict present-day values for $\Omega_{dust}$ which are in very good agreement with the estimate obtained
by \cite{Du11} in local galaxies.
On the other hand, the evolution of $\Omega_{dust}$ obtained in both DE and alternative scenarios are in reasonable agreement
with the evolution of the dust content in H-ATLAS galaxies, whereas the PLE scenario predicts a
weaker $\Omega_{dust}$ evolution than indicated by observations.\\
It is worth to stress that the overall satisfactory agreement between predictions and observations
is achieved without any further fine-tuning of the parameters of our models. 
One remarkable achievement is the coherency of the results obtained in the framework of
the alternative and DE scenarios developed here, which allows us to simultaneously
account for the evolution of the cosmic SFR and the cosmic dust budget in galaxies,
even though the latter claim is valid only at redshift $z<0.5$. \\
The PLE scenario predicts a much weaker evolution of the dust mass density in galaxies, 
and this is again coherent with the results discussed in Sect.~\ref{sec_CSFR}.
This also confirms that the behavior of galaxies
as traced by multi-wavelength observations at $z<0.5$ can only be explained with the assumption of an evolution of their number
density across the same redshift range. \\

In Fig.~\ref{fig:Omega_dust}, the $\Omega_{dust}$ estimates, except those of \cite{Du11}, are generally higher than our predictions.
This is mostly because these values also account for the amount of dust present outside galaxies,
i. e. in the intergalactic medium (IGM) or in galactic halos, hence they cannot be compared to our results,
which pertain to the galactic dust content only. 
Our models do not include prescriptions about dust processing outside the galaxy. 
An assessment of the dust budget outside galaxies will have to account for the amount
of dust and heavy elements ejected into the IGM throughout the cosmic history,
as well as for the additional creation and destruction processes which may occur in the intergalactic and intra-cluster media.

We also tested the case in which dust lost is taken into account, and
we found that the dust fraction lost by galactic wind represents a negligible fraction 
with respect to the total amount inside the galaxies.

One of the most  extended and comprehensive far-IR observational datasets of galaxies  is  the 
Herschel
Guaranteed Time Observation (GTO) PACS Evolutionary Probe (PEP) Survey \citep{Gr13}. 
By means of such a dataset, \citet{Gr13} studied the evolution of the far-IR galactic luminosity function up to $z\sim4$. 
In such a work, a galactic counterpart was found for most of the sources (>87$\%$), possibly leaving little room for 
extra-galactic emission in the far-IR and consequently to large reservoirs of 
dust outside galaxies. Currently, work is in progress in order to shed more light on this issue.


\begin{figure}
    \centering
      \includegraphics[width=0.48\textwidth]{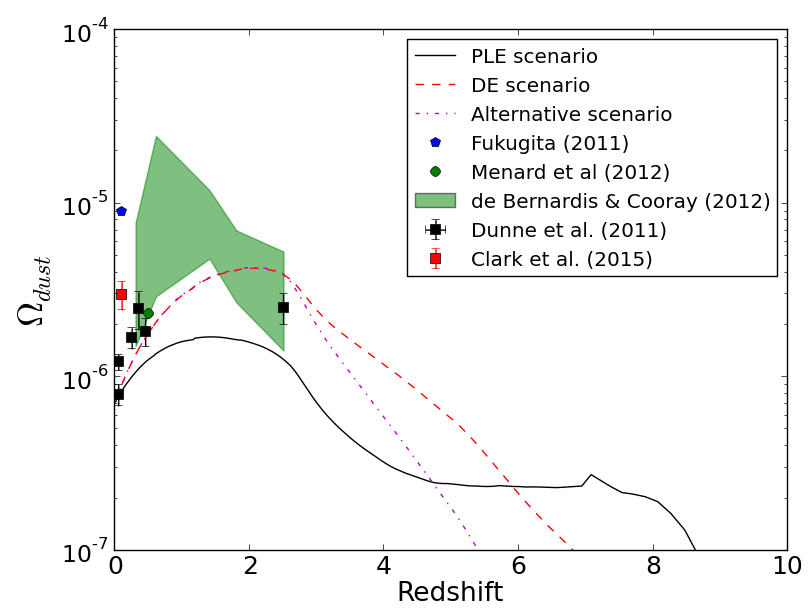}
       \caption{Evolution of $\Omega_{dust}$ as a function of the redshift in the 
       PLE (black solid), DE (red dashed) and alternative scenarios.}\label{fig:Omega_dust}
\end{figure}

\subsection{Evolution of the cosmic mean metallicity}

The cosmic mean metallicity (CMM) can be defined as 
\begin{equation}
  CMM =  \frac{\sum_k Z_{k} \cdot n_k}{\sum_k n_k},
  \label{eq:CMM}
 \end{equation}
(\citealt{EP97}, \citealt{Ca04}), where $Z_{k}$ is the average interstellar metallicity 
in galaxies of the $k-$th morphological type.
   
In Figure \ref{fig:CMM} we show the evolution of the cosmic mean metallicity defined in eq.~\ref{eq:CMM} predicted
by means of our models in the three scenarios considered in this work.
In each panel of Fig. \ref{fig:CMM} we show also the contributions of each morphological type, defined as  
$\frac{n_{k} Z_{k}}{n_{tot}}$, where $n_{tot} = \sum_k n_k$.  

Clearly, irrespective of the chosen scenario, the CMM increases as the redshift decreases. The global increase is only marginally dependent 
on the galaxy evolution scenario. \\
In the PLE scenario, the cosmic metallicity is dominated by the contribution of elliptical galaxies at very high redshift ($z>6$)
and by the contribution of spirals at $z<6$, with a negligible contribution from irregulars galaxies. 
This is the scenario characterized by the the steepest increase of the CMM, and this is clearly a consequence of 
the simultaneous, rapid formation of the entire population of elliptical galaxies, beginning at $z=10$.\\
In the DE scenario, the CMM is always completely dominated by the contribution of spirals. 
In this model, both irregulars and ellipticals give a negligible contribution to the cosmic mean metallicity at any redshift.\\
Once again, in the alternative scenario the evolution of the CMM is similar to the one of the DE scenario,  with the only difference of
a larger contribution from elliptical galaxies at redshift $z<4$. \\
Our results show that the increase of the CMM is relatively fast, and that on average, already 0.25 Gyr after the onset of 
star formation the interstellar matter has already been enriched to a value of $\sim~0.1$ solar. 
At $z=0$ all scenarios predict a solar CMM. This confirm previous results obtained in the past on the basis of an approach very similar to 
the one adopted here and based on galactic chemical evolution models (\citealt{Ca04}) or estimates based on an average of the metal 
abundances in various components (stars and interstellar matter) in various types of galaxies (\citealt{EP97}) and mass- and volume-averaged quantities 
computed from the integrated spectra of large databases, such as the Sloan Digital Sky Survey (\citealt{Ga08}).

Our estimate of the mean interstellar metallicity does not take into account the fraction of heavy elements subtracted 
from the gas phase and incorporated into dust grains. 
This fraction is commonly expressed by the dust-to-metals ratio, defined as the ratio between the total mass in dust $M_{dust}$ and 
the total interstellar mass in metals $M_{Z}$, i.e. 
\begin{equation}
 DTM =  \frac{M_{dust}}{M_z}
\label{eq:DTM}
\end{equation}
Our models for spirals and irregulars yield present-day dust-to-gas ratios  $DTM_{spi}= 0.68$ and $DTM_{irreg}= 0.60$. 
This implies that in the local Universe, the fraction of metals which will indeed be observable in the ISM represents $\sim~30-40\%$ of the total, hence 
cosmic mean 'undepleted' metallicity amounts to $\sim 0.005$ and $\sim 0.003$ in spirals and irregulars, respectively. 
Our estimate of the $DTM_{irreg}$ represents a typical value observed in low-metallicity systems analyzed by \citet{Dec13}.
Concerning elliptical galaxies, our estimate is $DTM_{ell}=0.039$, which is lower than in spirals and irregulars 
and in good agreement with \citet{Ca08}.

\begin{figure}
 \includegraphics[width=0.48\textwidth]{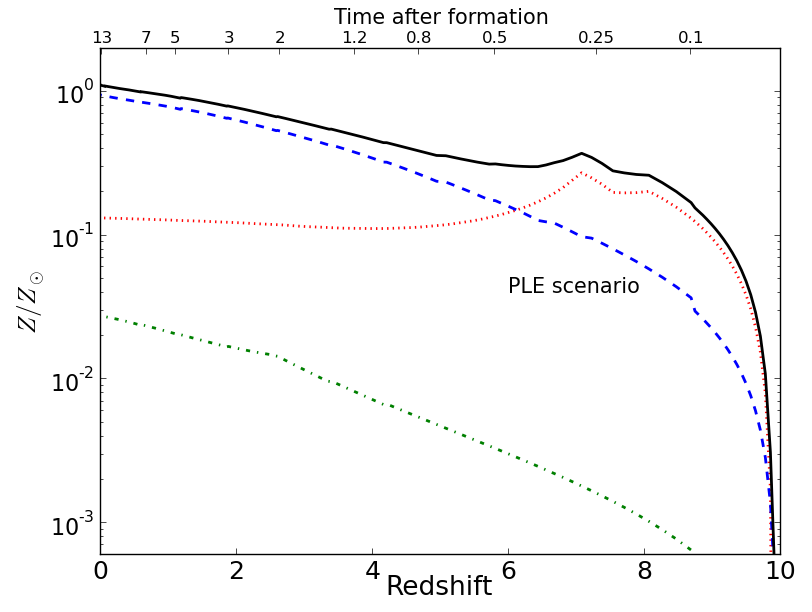}
 \includegraphics[width=0.48\textwidth]{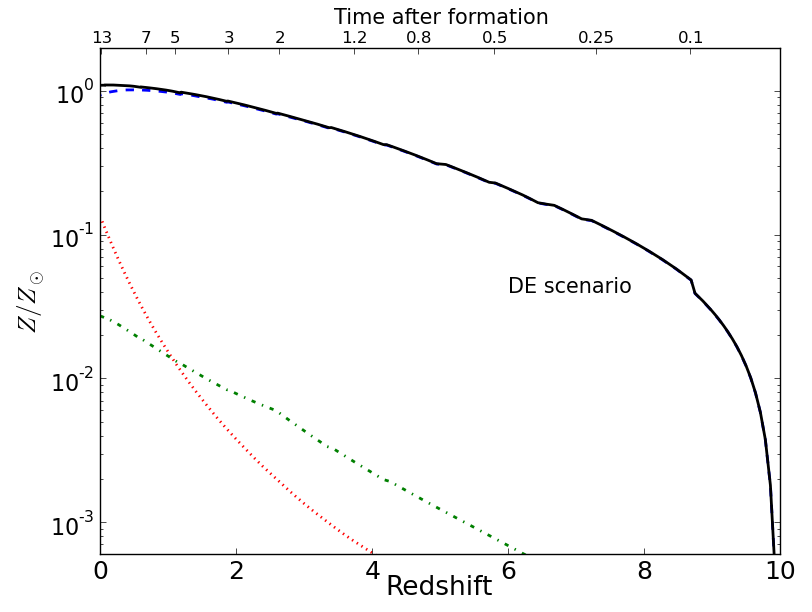}
 \includegraphics[width=0.48\textwidth]{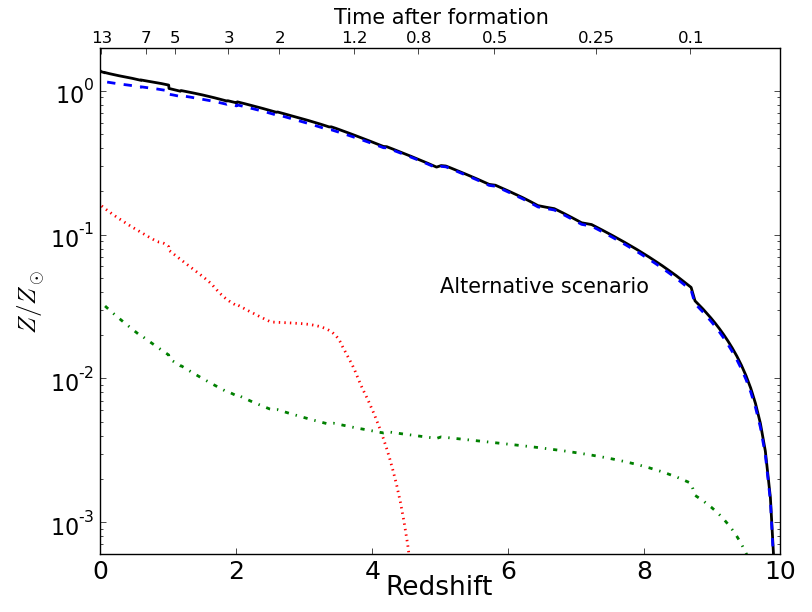}
 \caption{Metallicity evolution versus redshift for PLE (top panel), DE (middle panel) and 
 the alternative (bottom panel) scenarios. Red dotted, blue dashed and green dash-dotted lines refer 
 to the metallicity evolution of irregulars, spirals and ellipticals, respectively. 
 Black solid line represents the total amount of metallicity as the sum of the different morphological types.
 }\label{fig:CMM}
 \end{figure}

\section{Conclusions}\label{sec_conclusion}
In this work we have presented a study on cosmic
evolution of dust production, star formation and metallicity.
In order to calibrate the reference models of galaxies of different morphological type, 
we have compared our results with the amount of dust observed in irregular, spiral and elliptical galaxies.

Once we have constrained the models, we have considered three different cosmological scenarios of galaxy evolution:
(i) a PLE scenario, in which the number densities of spirals, ellipticals and irregulars are anchored to their present-day values 
and are not allowed to vary with redshift; 
(ii) a DE scenario, in which the number density is assumed to evolve strongly with redshift in order to mimic a hierarchical growth,
as suggested by the $\Lambda$CDM scenario; 
(iii) an alternative scenario, where ellipticals evolve in an observationally-motivated way. 
The main results of our paper can be summarized as follows: 
\begin{enumerate}
\item Our models of irregulars and spirals can reproduce the observed dust-to-gas ratio:
we found that irregular galaxies show a large spread of this quantity in the low metallicity range (7.5<$\log(O/H)+12$<8.0), whereas 
spirals have a narrower dispersion mainly concentrated at higher metallicities. 

\item From the study of the dust-to-gas ratios in different galaxies we found that dust production 
 by stars is the most important process at low metallicity, 
 whereas it plays a negligible role at high metallicity, in agreement with previous works (\citealt{Zh08}; \citealt{As13}).

\item The most important parameter which regulates the spread of the dust-to-gas ratio in irregular 
galaxies is the dust condensation of Type II SNe, which reflects the density of the ISM in which they explode. 
In general, the  denser the medium, the more resistance will be encountered by the SN shock and the more efficient will be the dust destruction.

\item Our model reproduces the large amount of dust observed in the high redshift counterparts of elliptical galaxies.
In principle, the dust crisis \citep{Ro14} can be solved without assuming a top-heavy IMF: an efficient contribution from Type II SNe beside dust growth 
is enough to explain the dust masses observed at high redshift. 
We have also shown that in order to avoid dust growth in such high redshift objects, 
much larger infall masses are needed. 
\item We predict the behavior of the dust production rate as a function of redshift in the three assumed galaxy formation scenarios.
In the DE and alternative scenario (based on the results of Pozzi et al. 2015), there is a peak in the cosmic dust rate in the redshift range $2<z<3$, 
while in the PLE scenario there is an additional peak at $z\sim 8$, due to the high redshift formation of ellipticals. 
\item Dust formation at high redshift, in each type of galaxy, is dominated by the contribution of massive stars and also by massive AGB stars.
Then, as the metallicity increases, also dust accretion plays a significant role.  
We assess the time-scale over which the bulk of the dust is formed in starburst galaxies. 
Even though this time-scale may change depending on the model parameters, 
in general in our models the bulk of the dust is formed within the first 0.25 Gyr.

\item In our analysis of the evolution of cosmic star formation, the scenario which provides the best agreement between model results and data 
is the alternative one, which allows us to reproduce the most up-to-date observations up to $z\sim 5$. 
On the other hand, within the DE scenario the agreement between model results and data is 
satisfactory only at $z<2$, whereas the PLE scenario appears completely ruled out by present data. 
While performing such an analysis, the reader has to be aware of the uncertainties affecting the CSFR data, 
especially in the high redshift Universe where the UV light should be corrected for dust extinction, and such a correction is in general highly uncertain \citep{Ma_Di14}.

\item We compute the evolution of the comoving interstellar dust density parameter $\Omega_{dust}$, finding a good agreement between our predictions and 
available data at $z<0.5$ within the DE and alternative scenarios. 
Generally, our estimates of $\Omega_{dust}$ are lower than other independent measurements. 
We found that the dust ejected by galactic winds is negligible with respect to the one formed inside galaxies. 
Therefore, the discrepancy between model predictions and observations could be attributed 
to a possible underestimation of our models of the dust 
lost in the intergalactic medium, 
or to a possible overestimation of the global amount of dust in some of the works discussed here (Section~\ref{sec_CDR}).


\item The global increase of the cosmic mean metallicity in galaxies with decreasing redshift is
only marginally dependent 
on the adopted galaxy evolution scenario. 
The PLE scenario predicts
a cosmic metallicity dominated by the contribution of ellipticals at very high redshift ($z>6$) , whereas spirals dominate at $z<6$. 
In the alternative scenario, the evolution of the CMM is similar to the one of the DE scenario,
with the difference of
a larger contribution from elliptical galaxies at redshift $z<4$. \\
The increase of the CMM is fast in every scenario: on average, already $\sim 0.25$ Gyr after the onset of 
star formation, the interstellar matter has already been enriched to a metallicity of $\sim~0.1$ solar.
At $z=0$ all scenarios predict a roughly solar CMM, confirming 
previous results based on galactic chemical evolution models (\citealt{Ca04}), averages of the metal 
abundances in various components (stars and interstellar matter) in various types of galaxies (\citealt{EP97}) and
observationally based estimates (\citealt{Ga08}).\\
In the local Universe, in star-forming galaxies a large fraction of heavy elements ($\sim 60-70 \%$)
is incorporated into solid grains, therefore not observable in the gas phase.

\end{enumerate}

\section*{Acknowledgements}
LG and FM acknowledge financial support from FRA2016 of the University of Trieste.
We also thank the anonymous referee for valuable suggestions that improved the clarity of the text.


\begin{thebibliography}{99}

\bibitem[Afonso et al.(2003)]{Afo03}   Afonso, J., Hopkins, A., Mobasher, B., Almeida, C. 2003, \apj, 597, 269

\bibitem[Asano et al.(2013)]{As13} Asano, R.~S., Takeuchi, T.~T., Hirashita, H., \& Nozawa, T.\ 2013, \mnras, 432, 637

\bibitem[Aoyama et al.(2017)]{Ao16} Aoyama, S., Hou, K.-C., Shimizu, I., et al.\ 2017, \mnras, 466, 105 


\bibitem[Bekki(2013)]{Be13} Bekki, K.\ 2013, \mnras, 432, 2298 

\bibitem[Bekki \& Tsujimoto(2014)]{Bek14} Bekki, K., \& Tsujimoto, T.\ 2014, \mnras, 444, 3879 

\bibitem[Bertoldi et al.(2003)]{Ber03} Bertoldi, F., Carilli, C.~L., Cox, P., et al.\ 2003, \aap, 406, L55 

\bibitem[Bradamante et al.(1998)]{Br98} Bradamante, F., Matteucci, F., \& D'Ercole, A.\ 1998, \aap, 337, 338 

\bibitem[Bianchi \& Schneider(2007)]{Bi_Sc_07} Bianchi, S., \& Schneider, R.\ 2007, \mnras, 378, 973 

\bibitem[Bocchio et al.(2014)]{Bo14} Bocchio, M., Jones, A.~P., \& Slavin, J.~D.\ 2014, \aap, 570, A32 

\bibitem[Bonaparte et al.(2013)]{Bon13} Bonaparte, I., Matteucci, F., Recchi, S., et al.\ 2013, \mnras, 435, 2460 

\bibitem[Calura et al.(2003)]{Ca03} Calura, F., Matteucci, F., \& Vladilo, G.\ 2003, \mnras, 340, 59 

\bibitem[Calura \& Matteucci(2003)]{CM03} Calura, F., Matteucci, F., 2003, ApJ, 596, 734

\bibitem[Calura \& Matteucci(2004)]{Ca04} Calura, F., Matteucci, F., 2004, MNRAS, MNRAS, 350, 351

\bibitem[Calura et al.(2007)]{Ca_Ma_To07} Calura, F., Matteucci, F., \& Tozzi, P.\ 2007, \mnras, 378, L11 

\bibitem[Calura et al.(2008)]{Ca08} Calura, F., Pipino, A., \& Matteucci, F.\ 2008, \aap, 479, 669

\bibitem[Calura et al.(2009)]{Ca09} Calura, F., Pipino, A., Chiappini, C., Matteucci, F., \& Maiolino, R.\ 2009, \aap, 504, 373 

\bibitem[Calura et al.(2014)]{Ca14} Calura, F., Gilli, R., Vignali, C., et al.\ 2014, \mnras, 438, 2765 

\bibitem[Calura et al.(2017)]{Ca17} Calura, F., Pozzi, F., Cresci, G., et al.\ 2017, \mnras, 465, 54 

\bibitem[Calzetti(2001)]{Ca01} Calzetti, D.\ 2001, \pasp, 113, 1449 

\bibitem[Carilli et al.(2004)]{Car04} Carilli, C.~L., Walter, F., Bertoldi, F., et al.\ 2004, \aj, 128, 997 

\bibitem[Casasola et al.(2017)]{Cas17} Casasola, V., Cassara, L.~P., Bianchi, S., et al.\ 2017, arXiv:1706.05351 

\bibitem[Chiappini et al.(1997)]{Ch97} Chiappini, C., Matteucci, F., \& Gratton, R.\ 1997, \apj, 477, 765 

\bibitem[Chiappini et al.(2001)]{Ch01} Chiappini, C., Matteucci, F., \& Romano, D.\ 2001, Galaxy Disks and Disk Galaxies, 230, 83  

\bibitem[Chomiuk \& Povich(2011)]{Ch_Po_11} Chomiuk, L., \& Povich, M.~S.\ 2011, \aj, 142, 197 

\bibitem[Clark et al.(2015)]{Cl15} Clark, C.~J.~R., Dunne, L., Gomez, H.~L., et al.\ 2015, \mnras, 452, 397 

\bibitem[Cooray et al.(2014)]{Co14} Cooray, A., Calanog, J., Wardlow, J.~L., et al.\ 2014, \apj, 790, 40 

\bibitem[Danziger et al.(1991)]{Da91} Danziger, I.~J., Bouchet, P., Gouiffes, C., \& Lucy, L.~B.\ 1991, The Magellanic Clouds, 148, 315 

\bibitem[De Bernardis \& Cooray(2012)]{De_Co12} De Bernardis, F., \& Cooray, A.\ 2012, \apj, 760, 14 

\bibitem[De Cia et al.(2013)]{Dec13} De Cia, A., Ledoux, C., Savaglio, S., Schady, P., \& Vreeswijk, P.~M.\ 2013, \aap, 560, A88 

\bibitem[De Cia et al.(2016)]{Dec16} De Cia, A., Ledoux, C., Mattsson, L., et al.\ 2016, \aap, 596, A97 

\bibitem[Desert et al.(1990)]{De90} Desert, F.-X., Boulanger, F., \& Puget, J.~L.\ 1990, \aap, 237, 215 

\bibitem[Draine \& Salpeter(1979)]{Dr_Sa_79} Draine, B.~T., \& Salpeter, E.~E.\ 1979, \apj, 231, 438 

\bibitem[Draine \& Lee(1984)]{Dr_Lee84} Draine, B.~T., \& Lee, H.~M.\ 1984, \apj, 285, 89 

\bibitem[Draine(1990)]{Dr90} Draine, B.~T.\ 1990, The Evolution of the Interstellar Medium, 12, 193.

\bibitem[Draine(2009)]{Dr09} Draine, B.~T.\ 2009, Cosmic Dust - Near and Far, 414, 453 

\bibitem[Dwek \& Scalo(1980)]{Dw_Sc_80} Dwek, E., \& Scalo, J.~M.\ 1980, \apj, 239, 193 

\bibitem[Dwek(1998)]{Dw98} Dwek, E.\ 1998, \apj, 501, 643 

\bibitem[Dwek et al.(2007)]{Dw07} Dwek, E., Galliano, F., \& Jones, A.~P.\ 2007, \apj, 662, 927 


\bibitem[Dunne et al.(2003)]{Du03} Dunne, L., Eales, S., Ivison, R., Morgan, H., \& Edmunds, M.\ 2003, \nat, 424, 285 


\bibitem[Dunne et al.(2011)]{Du11} Dunne, L., et al., 2011, MNRAS, 417, 1510

\bibitem[Eales et al.(2010)]{Ea10} Eales S., at al., 2010, PASP, 122, 499

\bibitem[Edmunds \& Phillips(1997)]{EP97} Edmunds, M. G., Phillips, S., 1997, MNRAS, 292, 733 

\bibitem[Fan et al.(2006)]{Fan06} Fan, X., Strauss, M.~A., Richards, G.~T., et al.\ 2006, \aj, 131, 1203 

\bibitem[Faucher-Gigu\'ere et al.(2008)]{Fau08} Faucher-Gigu\'ere, A., Lidz, A., Hernquist, L., Zaldarriaga, M., 2008, ApJL, 682, 9

\bibitem[Ferrara et al.(2016)]{Fe16} Ferrara, A., Viti, S., \& Ceccarelli, C.\ 2016, \mnras, 463, L112 

\bibitem[Ferrarotti \& Gail(2006)]{Fe_Ga_06} Ferrarotti, A.~S., \& Gail, H.-P.\ 2006, \aap, 447, 553 

\bibitem[Fukugita \& Peebles(2004)]{Fu_Pe04} Fukugita, M., \& Peebles, P.~J.~E.\ 2004, \apj, 616, 643 

\bibitem[Fukugita(2011)]{Fu11} Fukugita, M.\ 2011, arXiv:1103.4191 

\bibitem[Gail et al.(2009)]{Ga09} Gail, H.-P., Zhukovska, S.~V., Hoppe, P., \& Trieloff, M.\ 2009, \apj, 698, 1136 

\bibitem[Galametz et al.(2011)]{Gala11} Galametz, M., Madden, S.~C., Galliano, F., et al.\ 2011, \aap, 532, A56 

\bibitem[Gallazzi et al.(2008)]{Ga08} Gallazzi,A., Brinchmann, J., Charlot, S., White, S. D. M., 2008, MNRAS, 383, 1439

\bibitem[Gehrz(1989)]{Ge89} Gehrz, R.\ 1989, Interstellar Dust, 135, 445 

\bibitem[Gioannini et al.(2017)]{Gi17} Gioannini, L., Matteucci, F., Vladilo, G., \& Calura, F.\ 2017, \mnras, 464, 985 

\bibitem[Ginolfi et al.(2017)]{Gin17} Ginolfi, M., Graziani, L., Schneider, R., et al.\ 2017, arXiv:1707.05328 

\bibitem[Gomez et al.(2012)]{Go12} Gomez, H.~L., Krause, O., Barlow, M.~J., et al.\ 2012, \apj, 760, 96 

\bibitem[Granato et al.(2000)]{Granato2000} Granato, G.~L., Lacey, C.~G., Silva, L., et al.\ 2000, \apj, 542, 710  

\bibitem[Grebel(2004)]{Gre04} Grebel, E.~K.\ 2004, Origin and Evolution of the Elements, 234 

\bibitem[Grieco et al.(2012)]{Grieco12} Grieco, V., Matteucci, F., Meynet, G., et al.\ 2012, \mnras, 423, 3049 

\bibitem[Grieco et al.(2014)]{Grieco14} Grieco, V., Matteucci, F., Calura, F., et al.\ 2014, \mnras, 444, 1054 

\bibitem[Gruppioni et al.(2013)]{Gr13} Gruppioni, C., Pozzi, F., Rodighiero, G., et al.\ 2013, \mnras, 432, 23 

\bibitem[Habing(1996)]{Ha96} Habing, H.~J.\ 1996, \aapr, 7, 97 

\bibitem[Harris \& Zaritsky(2009)]{Ha_Za_09} Harris, J., \& Zaritsky, D.\ 2009, \aj, 138, 1243 

\bibitem[Hirashita(2000)]{Hi00} Hirashita, H.\ 2000, \pasj, 52, 585 

\bibitem[Hirashita \& Kuo(2011)]{Hi_Ku_11} Hirashita, H., \& Kuo, T.-M.\ 2011, \mnras, 416, 1340 

\bibitem[Hollenbach \& Salpeter(1971)]{Ho_Sa71} Hollenbach, D., \& Salpeter, E.~E.\ 1971, \apj, 163, 155 

\bibitem[Hopkins(2007)]{Ho07} Hopkins, A.~M.\ 2007, \apj, 654, 1175 

\bibitem[Indebetouw et al.(2014)]{In14} Indebetouw, R., Matsuura, M., Dwek, E., et al.\ 2014, \apjl, 782, L2 

\bibitem[Jenkins(2009)]{Je09} Jenkins, E.~B.\ 2009, \apj, 700, 1299 

\bibitem[Jones et al.(1994)]{Jo94} Jones, A.~P., Tielens, A.~G.~G.~M., Hollenbach, D.~J., \& McKee, C.~F.\ 1994, \apj, 433, 797 

\bibitem[Kennicutt(1998)]{Ke98} Kennicutt, R.~C., Jr.\ 1998, \araa, 36, 189 

\bibitem[Kennicutt et al.(2011)]{Ke11} Kennicutt, R.~C., Calzetti, D., Aniano, G., et al.\ 2011, \pasp, 123, 1347

\bibitem[Kennicutt \& Evans(2012)]{Ke_Ev12} Kennicutt, R.~C., \& Evans, N.~J.\ 2012, \araa, 50, 531 

\bibitem[Kistler et al.(2009)]{Ki09} Kistler, M.~D., Y{\"u}ksel, H., Beacom, J.~F., Hopkins, A.~M., \& Wyithe, J.~S.~B.\ 2009, \apjl, 705, L104 

\bibitem[Kuo \& Hirashita(2012)]{Kuo12} Kuo, T.-M., \& Hirashita, H.\ 2012, \mnras, 424, L34 

\bibitem[Laporte et al.(2017)]{Laporte17} Laporte, N., Ellis, R.~S., Boone, F., et al.\ 2017, \apjl, 837, L21 

\bibitem[Leroy et al.(2007)]{Le07} Leroy, A., Bolatto, A., Stanimirovic, S., et al.\ 2007, \apj, 658, 1027 

\bibitem[Liffman \& Clayton(1989)]{Li_Cl_89} Liffman, K., \& Clayton, D.~D.\ 1989, \apj, 340, 853 

\bibitem[Lucy et al.(1989)]{Lu89} Lucy, L.~B., Danziger, I.~J., Gouiffes, C., \& Bouchet, P.\ 1989, IAU Colloq.~120: Structure and Dynamics of the Interstellar Medium, 350, 164 

\bibitem[Madden et al.(2013)]{Ma13} Madden, S.~C., R{\'e}my-Ruyer, A., Galametz, M., et al.\ 2013, \pasp, 125, 600 

\bibitem[Madau \& Dickinson(2014)]{Ma_Di14} Madau, P., \& Dickinson, M.\ 2014, \araa, 52, 415 

\bibitem[Maiolino et al.(2015)]{Maiolino15} Maiolino, R., Carniani, S., Fontana, A., et al.\ 2015, \mnras, 452, 54 

\bibitem[Mancini et al.(2016)]{Mancini2016} Mancini, M., Schneider, R., Graziani, L., et al.\ 2016, \mnras, 462, 3130 

\bibitem[Mathis(1990)]{Ma90} Mathis, J.~S.\ 1990, \araa, 28, 37 

\bibitem[Matteucci \& Tornambe(1987)]{Ma_To87} Matteucci, F., \& Tornambe, A.\ 1987, \aap, 185, 51 

\bibitem[Matteucci(1994)]{Ma94} Matteucci, F.\ 1994, \aap, 288, 57 

\bibitem[Matteucci \& Recchi(2001)]{Ma_Re01} Matteucci, F., \& Recchi, S.\ 2001, \apj, 558, 351 


\bibitem[Matsuura et al.(2011)]{Ma11} Matsuura, M., Dwek, E., Meixner, M., et al.\ 2011, Science, 333, 1258 

\bibitem[Matsuura et al.(2015)]{Ma15} Matsuura, M., Dwek, E., Barlow, M.~J., et al.\ 2015, \apj, 800, 50 

\bibitem[Mattsson(2011)]{Matt11} Mattsson, L.\ 2011, \mnras, 414, 781 

\bibitem[Mattsson(2015)]{Matt15} Mattsson, L.\ 2015, arXiv:1505.04758 

\bibitem[McKinnon et al.(2016)]{McKinnon16} McKinnon, R., Torrey, P., \& Vogelsberger, M.\ 2016, \mnras, 457, 3775 

\bibitem[M{\'e}nard \& Fukugita(2012)]{Me_Fu12} M{\'e}nard, B., \& Fukugita, M.\ 2012, \apj, 754, 116 

\bibitem[Micali et al.(2013)]{Mi13} Micali, A., Matteucci, F., \& Romano, D.\ 2013, \mnras, 436, 1648 

\bibitem[Murante et al.(2015)]{Murante15} Murante, G., Monaco, P., Borgani, S., et al.\ 2015, \mnras, 447, 178 

\bibitem[Nanni et al.(2013)]{Na13} Nanni, A., Bressan, A., Marigo, P., \& Girardi, L.\ 2013, \mnras, 434, 2390 

\bibitem[Ota et al.(2014)]{Ota14} Ota, K., Walter, F., Ohta, K., et al.\ 2014, \apj, 792, 34 

\bibitem[Ouchi et al.(2013)]{Ou13}Ouchi M., et al., 2013, ApJ, 778, 102

\bibitem[Pei et al.(1991)]{Pei91} Pei, Y.~C., Fall, S.~M., \& Bechtold, J.\ 1991, \apj, 378, 6 

\bibitem[Pettini et al.(1994)]{Pe94} Pettini, M., Smith, L.~J., Hunstead, R.~W., \& King, D.~L.\ 1994, \apj, 426, 79 

\bibitem[Pipino \& Matteucci(2004)]{Pi_Ma04} Pipino, A., \& Matteucci, F.\ 2004, \mnras, 347, 968 

\bibitem[Pipino et al.(2011)]{Pip11} Pipino, A., Fan, X.~L., Matteucci, F., et al.\ 2011, \aap, 525, A61 

\bibitem[Piovan et al.(2011)]{Pi11} Piovan, L., Chiosi, C., Merlin, E., et al.\ 2011, arXiv:1107.4541 

\bibitem[Planck Collaboration et al.(2016)]{Pl16} Planck Collaboration, Ade, P.~A.~R., Aghanim, N., et al.\ 2016, \aap, 594, A13 

\bibitem[Popping et al.(2016)]{Po16} Popping, G., Somerville, R.~S., \& Galametz, M.\ 2016, arXiv:1609.08622 

\bibitem[Pozzi et al.(2015)]{Po15} Pozzi, F., Calura, F., Gruppioni, C., et al.\ 2015, \apj, 803, 35 

\bibitem[Recchi et al.(2002)]{Re_02} Recchi, S., Matteucci, F., \& D'Ercole, A.\ 2002, Chemical Enrichment of Intracluster and Intergalactic Medium, 253, 397 

\bibitem[R{\'e}my-Ruyer et al.(2014)]{Re14} R{\'e}my-Ruyer, A., Madden, S.~C., Galliano, F., et al.\ 2014, \aap, 563, A31 

\bibitem[R{\'e}my-Ruyer et al.(2015)]{Re15} R{\'e}my-Ruyer, A., Madden, S.~C., Galliano, F., et al.\ 2015, \aap, 582, A121 

\bibitem[Romano et al.(2005)]{Ro_05} Romano, D., Chiappini, C., Matteucci, F., \& Tosi, M.\ 2005, \aap, 430, 491 

\bibitem[Rowlands et al.(2014)]{Ro14} Rowlands, K., Gomez, H.~L., Dunne, L., et al.\ 2014, \mnras, 441, 1040 

\bibitem[Salpeter(1955)]{Sa95} Salpeter, E.~E.\ 1955, \apj, 121, 161 

\bibitem[Santini et al.(2014)]{Santini2014} Santini, P., Maiolino, R., Magnelli, B., et al.\ 2014, \aap, 562, A30 

\bibitem[Scalo(1986)]{Sc86} Scalo, J.~M.\ 1986, \fcp, 11, 1  

\bibitem[Schmidt(1959)]{Sc59} Schmidt, M.\ 1959, \apj, 129, 243 

\bibitem[Silva et al.(1998)]{Silva1998} Silva, L., Granato, G.~L., Bressan, A., \& Danese, L.\ 1998, \apj, 509, 103 

\bibitem[Spitoni et al.(2017)]{Spi17} Spitoni, E., Gioannini, L., \& Matteucci, F.\ 2017, In press., arXiv:1705.01297 

\bibitem[Todini \& Ferrara(2001)]{To_Fe_01} Todini, P., \& Ferrara, A.\ 2001, \mnras, 325, 726 

\bibitem[Valiante et al.(2009)]{Va09} Valiante, R., Schneider, R., Bianchi, S., \& Andersen, A.~C.\ 2009, \mnras, 397, 1661 

\bibitem[Valiante et al.(2014)]{Va14} Valiante, R., Schneider, R., Salvadori, S., \& Gallerani, S.\ 2014, \mnras, 444, 2442 

\bibitem[Ventura et al.(2012)]{Ve12} Ventura, P., Criscienzo, M.~D., Schneider, R., et al.\ 2012, \mnras, 424, 2345 

\bibitem[Ventura et al.(2014)]{Ve14} Ventura, P., Dell'Agli, F., Schneider, R., et al.\ 2014, \mnras, 439, 977 

\bibitem[Vincoletto et al.(2012)]{Vincoletto12} Vincoletto, L., Matteucci, F., Calura, F., Silva, L., \& Granato, G.\ 2012, \mnras, 421, 3116 

\bibitem[Vladilo(2002)]{Vla02} Vladilo, G.\ 2002, \aap, 391, 407 

\bibitem[Vladilo(2004)]{Vl04} Vladilo, G.\ 2004, \aap, 421, 479 

\bibitem[Vladilo \& P{\'e}roux(2005)]{Vl_Pe05} Vladilo, G., \& P{\'e}roux, C.\ 2005, \aap, 444, 461 

\bibitem[Watson et al.(2015)]{Wa15} Watson, D., Christensen, L., Knudsen, K.~K., et al.\ 2015, \nat, 519, 327 

\bibitem[Whelan \& Iben(1973)]{Wh_Ib_73} Whelan, J., \& Iben, I., Jr.\ 1973, \apj, 186, 1007 

\bibitem[Willott et al.(2010)]{Wi10} Willott, C.~J., Delorme, P., Reyl{\'e}, C., et al.\ 2010, \aj, 139, 906 

\bibitem[Witt \& Gordon(2000)]{Wi_Go00} Witt, A.~N., \& Gordon, K.~D.\ 2000, \apj, 528, 799 

\bibitem[Zavala et al.(2015)]{Za15} Zavala, J.~A., Micha{\l}owski, M.~J., Aretxaga, I., et al.\ 2015, \mnras, 453, L88 


\bibitem[Zhukovska et al.(2008)]{Zh08} Zhukovska, S., Gail, H.-P., \& Trieloff, M.\ 2008, \aap, 479, 453 

\bibitem[Zhukovska(2014)]{Zhu14} Zhukovska, S.\ 2014, \aap, 562, A76 

\bibitem[Zhukovska et al.(2016)]{Zhu16} Zhukovska, S., Dobbs, C., Jenkins, E.~B., \& Klessen, R.~S.\ 2016, \apj, 831, 147 

\bibitem[Zubko et al.(2004)]{Zu04} Zubko, V., Dwek, E., \& Arendt, R.~G.\ 2004, \apjs, 152, 211 
 
\end{thebibliography}
\end{document}